\documentclass[aps,prd,reprint,longbibliography,superscriptaddress,nofootinbib]{revtex4-2}

\usepackage{gLEEStyle}

\begin{document}
\title{
Measurement of neutral current single $\pi^0$ production on argon with the MicroBooNE detector}

\newcommand{\Bern}{Universit{\"a}t Bern, Bern CH-3012, Switzerland}
\newcommand{\BNL}{Brookhaven National Laboratory (BNL), Upton, NY, 11973, USA}
\newcommand{\UCSB}{University of California, Santa Barbara, CA, 93106, USA}
\newcommand{\Cambridge}{University of Cambridge, Cambridge CB3 0HE, United Kingdom}
\newcommand{\CIEMAT}{Centro de Investigaciones Energ\'{e}ticas, Medioambientales y Tecnol\'{o}gicas (CIEMAT), Madrid E-28040, Spain}
\newcommand{\Chicago}{University of Chicago, Chicago, IL, 60637, USA}
\newcommand{\Cincinnati}{University of Cincinnati, Cincinnati, OH, 45221, USA}
\newcommand{\CSU}{Colorado State University, Fort Collins, CO, 80523, USA}
\newcommand{\Columbia}{Columbia University, New York, NY, 10027, USA}
\newcommand{\Edinburgh}{University of Edinburgh, Edinburgh EH9 3FD, United Kingdom}
\newcommand{\FNAL}{Fermi National Accelerator Laboratory (FNAL), Batavia, IL 60510, USA}
\newcommand{\Granada}{Universidad de Granada, Granada E-18071, Spain}
\newcommand{\Harvard}{Harvard University, Cambridge, MA 02138, USA}
\newcommand{\IIT}{Illinois Institute of Technology (IIT), Chicago, IL 60616, USA}
\newcommand{\KSU}{Kansas State University (KSU), Manhattan, KS, 66506, USA}
\newcommand{\Lancaster}{Lancaster University, Lancaster LA1 4YW, United Kingdom}
\newcommand{\LANL}{Los Alamos National Laboratory (LANL), Los Alamos, NM, 87545, USA}
\newcommand{\Louisiana}{Louisiana State University, Baton Rouge, LA, 70803, USA}
\newcommand{\Manchester}{The University of Manchester, Manchester M13 9PL, United Kingdom}
\newcommand{\MIT}{Massachusetts Institute of Technology (MIT), Cambridge, MA, 02139, USA}
\newcommand{\Michigan}{University of Michigan, Ann Arbor, MI, 48109, USA}
\newcommand{\Minnesota}{University of Minnesota, Minneapolis, MN, 55455, USA}
\newcommand{\NMSU}{New Mexico State University (NMSU), Las Cruces, NM, 88003, USA}
\newcommand{\Oxford}{University of Oxford, Oxford OX1 3RH, United Kingdom}
\newcommand{\Pitt}{University of Pittsburgh, Pittsburgh, PA, 15260, USA}
\newcommand{\Rutgers}{Rutgers University, Piscataway, NJ, 08854, USA}
\newcommand{\SLAC}{SLAC National Accelerator Laboratory, Menlo Park, CA, 94025, USA}
\newcommand{\SDSMT}{South Dakota School of Mines and Technology (SDSMT), Rapid City, SD, 57701, USA}
\newcommand{\Maine}{University of Southern Maine, Portland, ME, 04104, USA}
\newcommand{\Syracuse}{Syracuse University, Syracuse, NY, 13244, USA}
\newcommand{\TelAviv}{Tel Aviv University, Tel Aviv, Israel, 69978}
\newcommand{\Tennessee}{University of Tennessee, Knoxville, TN, 37996, USA}
\newcommand{\UTA}{University of Texas, Arlington, TX, 76019, USA}
\newcommand{\Tufts}{Tufts University, Medford, MA, 02155, USA}
\newcommand{\VTech}{Center for Neutrino Physics, Virginia Tech, Blacksburg, VA, 24061, USA}
\newcommand{\Warwick}{University of Warwick, Coventry CV4 7AL, United Kingdom}
\newcommand{\Yale}{Wright Laboratory, Department of Physics, Yale University, New Haven, CT, 06520, USA}

\affiliation{\Bern}
\affiliation{\BNL}
\affiliation{\UCSB}
\affiliation{\Cambridge}
\affiliation{\CIEMAT}
\affiliation{\Chicago}
\affiliation{\Cincinnati}
\affiliation{\CSU}
\affiliation{\Columbia}
\affiliation{\Edinburgh}
\affiliation{\FNAL}
\affiliation{\Granada}
\affiliation{\Harvard}
\affiliation{\IIT}
\affiliation{\KSU}
\affiliation{\Lancaster}
\affiliation{\LANL}
\affiliation{\Louisiana}
\affiliation{\Manchester}
\affiliation{\MIT}
\affiliation{\Michigan}
\affiliation{\Minnesota}
\affiliation{\NMSU}
\affiliation{\Oxford}
\affiliation{\Pitt}
\affiliation{\Rutgers}
\affiliation{\SLAC}
\affiliation{\SDSMT}
\affiliation{\Maine}
\affiliation{\Syracuse}
\affiliation{\TelAviv}
\affiliation{\Tennessee}
\affiliation{\UTA}
\affiliation{\Tufts}
\affiliation{\VTech}
\affiliation{\Warwick}
\affiliation{\Yale}

\author{P.~Abratenko} \affiliation{\Tufts}
\author{J.~Anthony} \affiliation{\Cambridge}
\author{L.~Arellano} \affiliation{\Manchester}
\author{J.~Asaadi} \affiliation{\UTA}
\author{A.~Ashkenazi}\affiliation{\TelAviv}
\author{S.~Balasubramanian}\affiliation{\FNAL}
\author{B.~Baller} \affiliation{\FNAL}
\author{C.~Barnes} \affiliation{\Michigan}
\author{G.~Barr} \affiliation{\Oxford}
\author{J.~Barrow} \affiliation{\MIT}\affiliation{\TelAviv}
% ...

\collaboration{The MicroBooNE Collaboration}
\thanks{microboone\_info@fnal.gov}\noaffiliation

\date{\today}
\begin{abstract}
We report the first measurement of $\pi^0$ production in neutral current (NC) interactions on argon with average neutrino energy of $\lesssim1$~GeV. We use data from the MicroBooNE detector's 85-tonne active volume liquid argon time projection chamber situated in Fermilab's Booster Neutrino Beam and exposed to $5.89\times10^{20}$ protons on target for this measurement. Measurements of NC $\pi^0$ events are reported for two exclusive event topologies without charged pions. Those include a topology with two photons from the decay of the $\pi^0$ and one proton and a topology with two photons and zero protons. Flux-averaged cross-sections for each exclusive topology and for their semi-inclusive combination are extracted (efficiency-correcting for two-plus proton final states), and the results are compared to predictions from the \textsc{genie}, \textsc{neut}, and \textsc{NuWro} neutrino event generators. We measure cross sections of $1.243\pm0.185$ (syst) $\pm0.076$ (stat), $0.444\pm0.098\pm0.047$, and $0.624\pm0.131\pm0.075$ $[10^{-38}\textrm{cm}^2/\textrm{Ar}]$ for the semi-inclusive NC$\pi^0$, exclusive NC$\pi^0$+1p, and exclusive NC$\pi^0$+0p processes, respectively. 

\end{abstract}
\maketitle


\newpage

\section{Introduction}

Neutrino-nucleus cross-sections have been the subject of intense study both experimentally and within the theory community in recent years due to their role in interpreting neutrino oscillation measurements and searches for other rare processes in neutrino scattering \cite{Benhar:2015wva}. While neutrino oscillation experiments primarily rely on measuring the rate of charged current (CC) interactions, it is also important that we build a solid understanding of inclusive and exclusive neutral current (NC) neutrino interactions.

NC neutrino interactions are of particular importance to $\nu_e$ and \nuebar{} measurements in the energy range of a few hundred MeV. This is especially true for detectors that cannot perfectly differentiate between photon- and electron-induced electromagnetic showers, and therefore where NC $\pi^0$ production followed by the subsequent decay $\pi^0 \rightarrow \gamma \gamma$ can be misidentified as $\nu_e$ or \nuebar{} CC scattering. Misidentification of photons as electrons complicates the interpretation of $\nu_e$ appearance measurements aiming to measure subtle signals. These include ~sterile neutrino oscillation searches with the upcoming Short Baseline Neutrino (SBN) experimental program \cite{SBNproposal} and CP violation measurements and mass hierarchy determination with the future Deep Underground Neutrino Experiment (DUNE) \cite{DUNE:2020ypp}. 

Furthermore, NC $\pi^0$ events can contribute as background to searches for rare neutrino scattering processes such as NC $\Delta$ resonance production followed by $\Delta$ radiative decay, or NC coherent single-photon production at energies below 1~GeV \cite{MicroBooNE:2021zai}. This is primarily a consequence of the limited geometric acceptance of some detectors, whereby one of the photons from a $\pi^0$ decay can escape the active volume of the detector. Depending on a detector's ability to resolve electromagnetic shower substructure, NC $\pi^0$ events can further contribute as background to searches for new physics beyond the Standard Model (BSM), such as $e^+e^-$ production predicted by a number of BSM models \cite{Bertuzzo:2018itn,Ballett:2018ynz,Abdullahi:2020nyr,Dutta:2020scq,Abdallah:2020vgg}.

Finally, NC measurements themselves can provide a unique channel for probing new physics. For example, searches for non-unitarity in the three-neutrino paradigm or searches for active to sterile neutrino oscillations are possible via NC rate disappearance measurements \cite{Cianci:2017okw,Furmanski:2020smg}. Such searches can provide complementary information to non-unitarity or light sterile neutrino oscillation parameters otherwise accessible only through CC measurements. 

Using a liquid argon time projection chamber (LArTPC) as its active detector, MicroBooNE \cite{MicroBooNE:2016pwy} shares the same technology and neutrino target nucleus as the upcoming SBN and future DUNE experiments. MicroBooNE's 85 metric ton active volume LArTPC is situated 468.5~m away from the proton beam target in the muon-neutrino-dominated Booster Neutrino Beam (BNB) at Fermilab~\cite{BNBflux} which is also used by SBN. The resulting neutrino beam has a mean energy $\langle E_\nu\rangle = 0.8$~GeV and is composed of $93.7\%~\nu_\mu$, $5.8\%~\bar{\nu}_\mu$, and $0.5\%~\nu_e/\bar{\nu}_e$. MicroBooNE's cross-section measurements on argon are therefore timely and directly relevant to these future (SBN and DUNE) programs.

We present the first measurement of neutrino-induced NC single-$\pi^0$ ($1\pi^0$) production on argon with a mean neutrino energy in the 1~GeV regime, which is also the highest-statistics measurement of this interaction channel on argon to date. This measurement is relevant to the physics programs of experiments that operate in the few-GeV regime (SBN~\cite{SBNproposal}, DUNE~\cite{DUNE:2020ypp}, NO$\nu$A~\cite{MINERvA:2016iqn,NOvA:2019bdw}, T2K~\cite{T2Kflux}, and Hyper-K~\cite{Hyper-KamiokandeProto-:2015xww}), especially those which share argon as a target material. Additionally, this measurement has been used to provide an indirect constraint to the rate of NC $1\pi^0$ backgrounds in MicroBooNE's recent search for a single-photon excess \cite{MicroBooNE:2021zai}. The only previous results for NC $1\pi^0$ scattering on argon are from the ArgoNeuT collaboration using the NuMI beam which has a much higher mean neutrino beam energy of 9.6 GeV for $\nu_\mu$ and of 3.6 GeV for $\overline{\nu}_\mu$  \cite{ArgoNeuT:2015ldo}. 

The interaction final states that are measured in this analysis are defined as
\begin{eqnarray}
\label{reaction}
\nu + \mathcal{A} \rightarrow \nu + \mathcal{A'} + \pi^0 + X,
\end{eqnarray}
where $\mathcal{A}$ represents the struck (argon) nucleus, $\mathcal{A'}$ represents the residual nucleus, and $X$ represents exactly one or zero protons plus any number of neutrons, but no other hadrons or leptons. The protons are identifiable in the MicroBooNE LArTPC by their distinct ionizing tracks while the $\pi^0$ is identifiable through the presence of two distinct electromagnetic showers, one for each photon from the $\pi^0\rightarrow\gamma\gamma$ decay, with kinematic properties such that they reconstruct to approximately the $\pi^0$ invariant mass.  

These one proton and zero proton samples are used first to perform a rate validation check and subsequently in three distinct cross-section measurements. By leveraging the capability of LArTPCs to detect and identify protons we perform the world's first exclusive NC$\pi^0$+0p and NC$\pi^0$+1p cross-section extractions and additionally measure the cross-section for NC$\pi^0$ interactions semi-inclusively using both the one proton and zero proton samples combined. Each of these cross-section extractions utilizes a distinct signal definition. The signal definitions for the two exclusive measurements place a threshold on true proton kinetic energy of greater than 50 MeV, while the semi-inclusive measurement allows for any number of protons. The signal definitions for all three measurements also require that there are no other hadrons or leptons in the final state (as noted above). MeV-scale photons, which may arise from nuclear de-excitation processes within the struck nucleus, are allowed in the final state. Finally, the signal definitions allow for interactions of all flavors of neutrinos that are present: $\nu_\mu$, $\bar{\nu}_\mu$, $\nu_e$, and $\bar{\nu}_e$.

These definitions are comparable to other historical NC $\pi^0$ measurements which typically require one and only one $\pi^0$ meson and little hadronic activity in the detector \cite{Barish:1974fe,Derrick:1980xw,Lee:1976wr,GargamelleNeutrinoPropane:1977hya,MiniBooNE:2008mmr,MiniBooNE2011,MiniBooNE:2009dxl,SciBooNE:2010NCpi0,K2K:2004qpv,T2K:2017epu}. This differs from the more inclusive approach of the ArgoNeuT experiment motivated both by its higher energy beam as well as the need to mitigate the low statistics of its data sample \cite{ArgoNeuT:2015ldo}. Making use of the MicroBooNE LArTPC's power in examining hadronic final state multiplicities and kinematic properties with high resolution, the flux-averaged cross-sections extracted in this analysis extend our understanding of this important interaction channel. The exclusive cross-sections reported provide new information useful for the tuning of NC $1\pi^0$ production and nuclear final state interactions in neutrino-argon scattering models, while the semi-inclusive cross-section enables a direct comparison to the MiniBooNE measurement of NC $\pi^0$ production.
 
\section{Analysis Overview}

This measurement uses data corresponding to a BNB exposure of $5.89\times10^{20}$~protons on target (POT), collected during the period 2016--2018 and referred to as ``Runs 1--3'' in many of the subsequent figures. Neutrino-argon interactions are simulated using a custom tune \cite{MicroBooNE:2021ccs} of the \textsc{genie} neutrino event generator v3.0.6 \cite{Andreopoulos:2009rq,GENIE:2021zuu} (based on model set  G18\_10a\_02\_11a) adopted by the MicroBooNE Collaboration. This tune specifically targets CC quasi-elastic (QE) and CC multi-nucleon interaction models and overall has very little direct effect on this NC-focused analysis. \textsc{genie} v3 uses the Berger-Sehgal~\cite{Rein:1980wg,Berger:2007rq} model for resonant production of $\pi^0$ and includes improved agreement with an expanded data set for the $A$-dependence of final state interactions (FSI), updated form factors \cite{Graczyk:2008}, updated diagrams for pion production processes \cite{Berger:2007rq,Kuzmin:2004,Nowak:2009}, and a new tune to neutrino-proton and neutrino-deuterium cross-section data \cite{GENIE:2021zuu}. The MicroBooNE Monte Carlo (MC) prediction further makes use of \textsc{geant4} v4\_10\_3\_03c \cite{GEANT4:2002zbu} for particle propagation and re-interactions within the detector and a custom detector response model all implemented within the LArSoft framework~\cite{Snider_2017}. 

The MicroBooNE data and MC reconstruction chain begins by reading out and processing the ionization charge signals detected on the 8,192 wires that make up the three anode planes of the MicroBooNE LArTPC. The procedure includes noise removal \cite{MicroBooNE:2017qiu} and signal processing as described in \cite{MicroBooNE:2018swd} and \cite{MicroBooNE:2018vro}. Localized regions of interest referred to as ``hits'' are then identified and fit to Gaussian pulses. The collection of these hits and their characteristics such as readout time, wire channel number, and integrated charge are then used as input to the Pandora pattern recognition framework for further processing \cite{Marshall:2015rfa}. The Pandora framework clusters and matches hits across three 2D projected views of the MicroBooNE active TPC volume to form 3D reconstructed objects. These objects are then classified as track-like or shower-like based on a multivariate classifier score and aggregated into candidate neutrino interactions. Pandora also reconstructs a candidate neutrino interaction vertex based on the position and orientation of the reconstructed tracks and showers which represents the most likely position of the neutrino interaction.

Being a surface detector, MicroBooNE is subject to a constant stream of high-energy cosmic rays impinging on the detector that substantially outnumber the neutrino interactions and form the largest background to candidate neutrino interactions. To incorporate the effect of cosmic-ray contamination in the simulation, cosmic ray data recorded \textit{in situ} at MicroBooNE, when the beam is not present, are used as overlays (at the wire signal waveform level) to  simulated neutrino interactions. During the 2.3 ms that it takes to ``drift'' ionization charge associated with neutrino interaction final states across the maximum 2.56 m drift distance, $\mathcal{O}(10)$ cosmic rays are expected to enter the detector. In order to reduce this cosmic-ray contamination, scintillation light recorded by the MicroBooNE photo-detector system is matched to candidate neutrino interactions during reconstruction and is also required to occur in time with the 1.6~$\mu$s long BNB neutrino spill.  

To select a high-purity sample of BNB neutrino NC $1\pi^0$ interactions, a series of topological, pre-selection, and boosted decision tree (BDT)-based selections are applied. This results in two mutually exclusive final selection topologies: $2\gamma1p$, which targets two photons and one proton in the final state, and $2\gamma0p$, which targets two photons and zero protons in the final state. The different selection stages are described below, along with the details of the systematic uncertainty evaluation. 

\subsection{Topological Selection and Pre-Selection}

The event selection begins with topology-based criteria for candidate neutrino interactions identified by Pandora and targets two mutually exclusive topological definitions: (a) two showers and one track ($2\gamma 1p$), and (b) two showers and zero tracks ($2\gamma 0p$). The two showers correspond to the photons expected from $\pi^0$ decay. The presence of a track corresponds to a reconstructed proton exiting the nucleus while the zero-track case suggests either a low-energy proton that is not reconstructed or no charged hadrons at all exiting the nucleus. 

Once events with the desired signal topologies are identified, a series of loose ``pre-selection'' requirements is applied to reduce obvious backgrounds or mis-reconstructed events. These pre-selection requirements include shower energy thresholds of 30~MeV for the leading shower and 20~MeV for the subleading shower in both topologies. The pre-selection also requires that the reconstructed neutrino interaction point be contained in a fiducial volume, defined as at least 5~cm away from any TPC wall, in order to help reduce the number of selected events with tracks that exit the detector. For the 2$\gamma$1p topology, the non-zero conversion distance of photons is explicitly used by requiring that each shower has a reconstructed start point of at least 1~cm from the reconstructed neutrino interaction vertex. Typically the reconstructed neutrino interaction vertex is identified as the start of the reconstructed proton candidate track. In order to remove a very small number of poorly reconstructed events in which the candidate track is not consistent with the hypothesis of originating from the candidate neutrino interaction vertex, a requirement is placed to ensure the track start point is always within 10~cm of the reconstructed neutrino interaction vertex. The efficiency of selecting NC $1\pi^0$ + 0 (1)p events using these pre-selection requirements is 21.5\% (19.9\%). Note that the efficiency of the 1p selection is lower because of the additional requirements placed on the track reconstruction. 

\subsection{Boosted Decision Tree-Based Selection}
After applying the pre-selection requirements, the remaining signal and background are further differentiated and separated using two tailored BDTs trained on simulation. The gradient boosting algorithm XGBoost \cite{Chen:2016:XST:2939672.2939785} is used to train each of the BDTs. They take as input various reconstructed kinematic, geometric, and calorimetric variables both for the signal (defined as an NC interaction with identically one $\pi^0$ in the final state) and for the background interactions. Because the two tailored BDTs target different topologies, notably including one track in the case of $2\gamma1p$ and zero tracks in the case of $2\gamma0p$, the signal definitions used for the two BDTs are slightly different. NC $\pi^0$ events with exactly one proton with true kinetic energy above 20~MeV are used as the training signal for the $2\gamma1p$ BDT while NC $\pi^0$ events with no protons with true kinetic energy above 20 MeV are used as the training signal for the $2\gamma0p$ BDT. We note that the 20 MeV threshold used in the BDT training is lower than the 50 MeV proton kinetic energy threshold used later during cross-section extraction, as during training we are aiming to push the threshold as low as possible. Each BDT is trained on ten reconstructed variables. Due to the existence of a proton candidate track in the $2\gamma1p$ sample, these ten variables differ for each BDT. They are listed below.\\

\noindent \textbf{Variables used in both $\bm{2\gamma1p}$ and $ \bm{2\gamma0p}$ BDTs:}
\begin{itemize}
    \item Leading and subleading shower impact parameters: The perpendicular distance between the back-projection of the reconstructed shower and the candidate neutrino interaction point which is a metric of how well each shower ``points'' back to the reconstructed neutrino interaction point. 
    \item Leading and subleading shower conversion distances: Defined as the distance between the reconstructed start of the shower and reconstructed neutrino interaction point. 
    \item Reconstructed energy of the leading shower.
\end{itemize}
\noindent \textbf{Variables used in only the $\bm{2\gamma1p}$ BDT:}
\begin{itemize}
    \item Reconstructed track length.
    \item Reconstructed track vertical angle: Defined as the arctangent of the track direction in the vertical plane with respect to the beam axis.
    \item Distance from track end to TPC wall: Calculated as the shortest distance to the closest TPC wall.
    \item Reconstructed mean energy deposition per unit length ($dE/dx$) of the track.
    \item Ratio of $dE/dx$ of the first half of track to that of the second half of the track: A metric for identifying stopping proton tracks that contain a Bragg peak.
\end{itemize}
\noindent \textbf{Variables used in only the $\bm{2\gamma0p}$ BDT:}
\begin{itemize}
    \item Reconstructed energy of the subleading shower.
    \item Leading and subleading shower geometric length per unit energy: The ratio of each shower's geometric length to its reconstructed energy. The geometric length is an estimate of the 3D extent of the electromagnetic shower. 
    \item Pandora ``neutrino score'': A multivariate classifier in the Pandora reconstruction suite which scores all reconstructed neutrino candidates based on their geometric and kinematic features as to how likely a candidate is due to a neutrino interaction or cosmic in origin.
    \item Reconstructed leading shower vertical angle: Direction in the vertical plane with respect to the beam axis.
\end{itemize}

\begin{figure}[h]
    \centering 
    \begin{subfigure}{0.49\textwidth}
        \includegraphics[width = \textwidth]{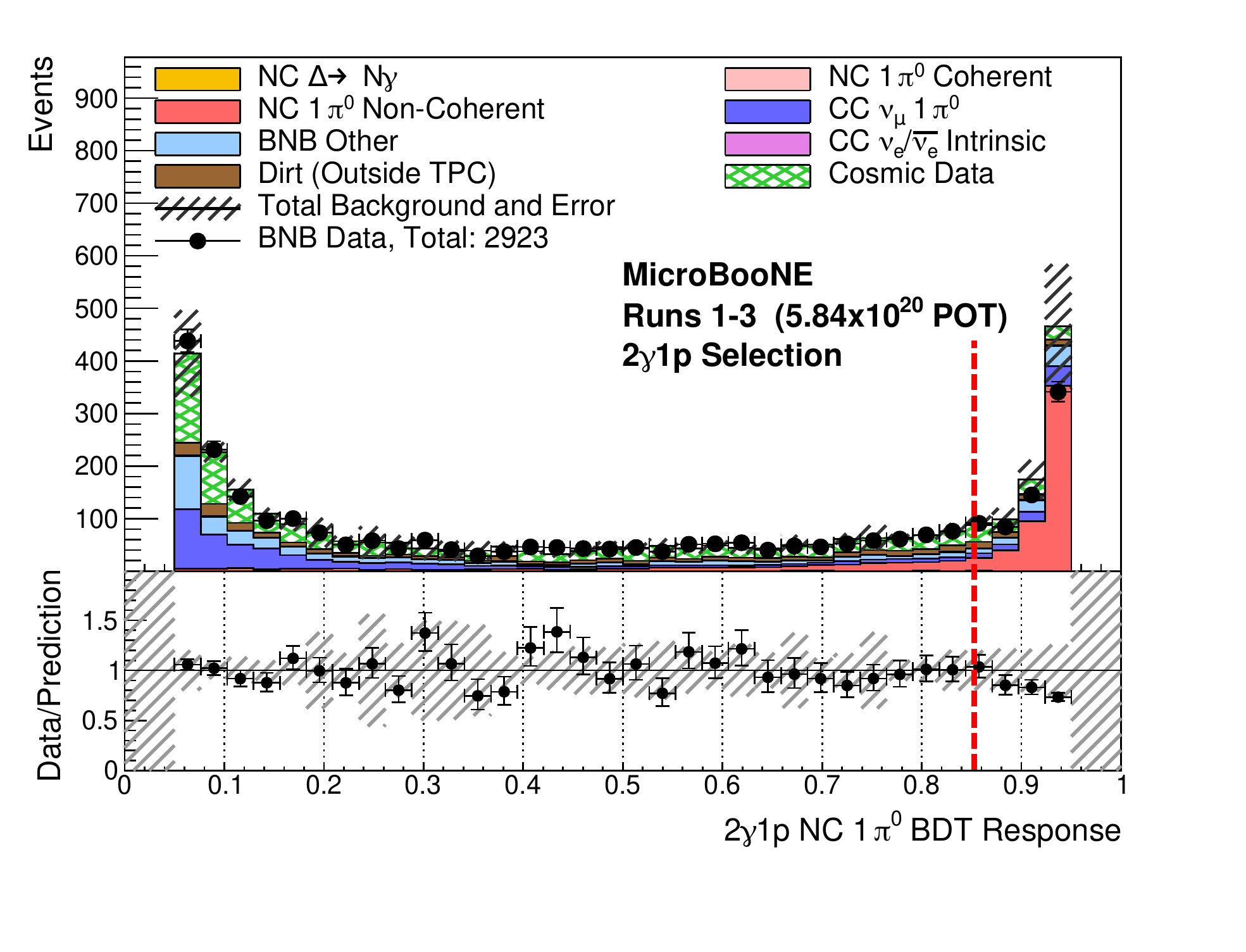}
        \vspace{-1cm}
        \caption{$2\gamma 1p$}
  \end{subfigure} 
  \begin{subfigure}{0.49\textwidth}
      \includegraphics[width = \textwidth]{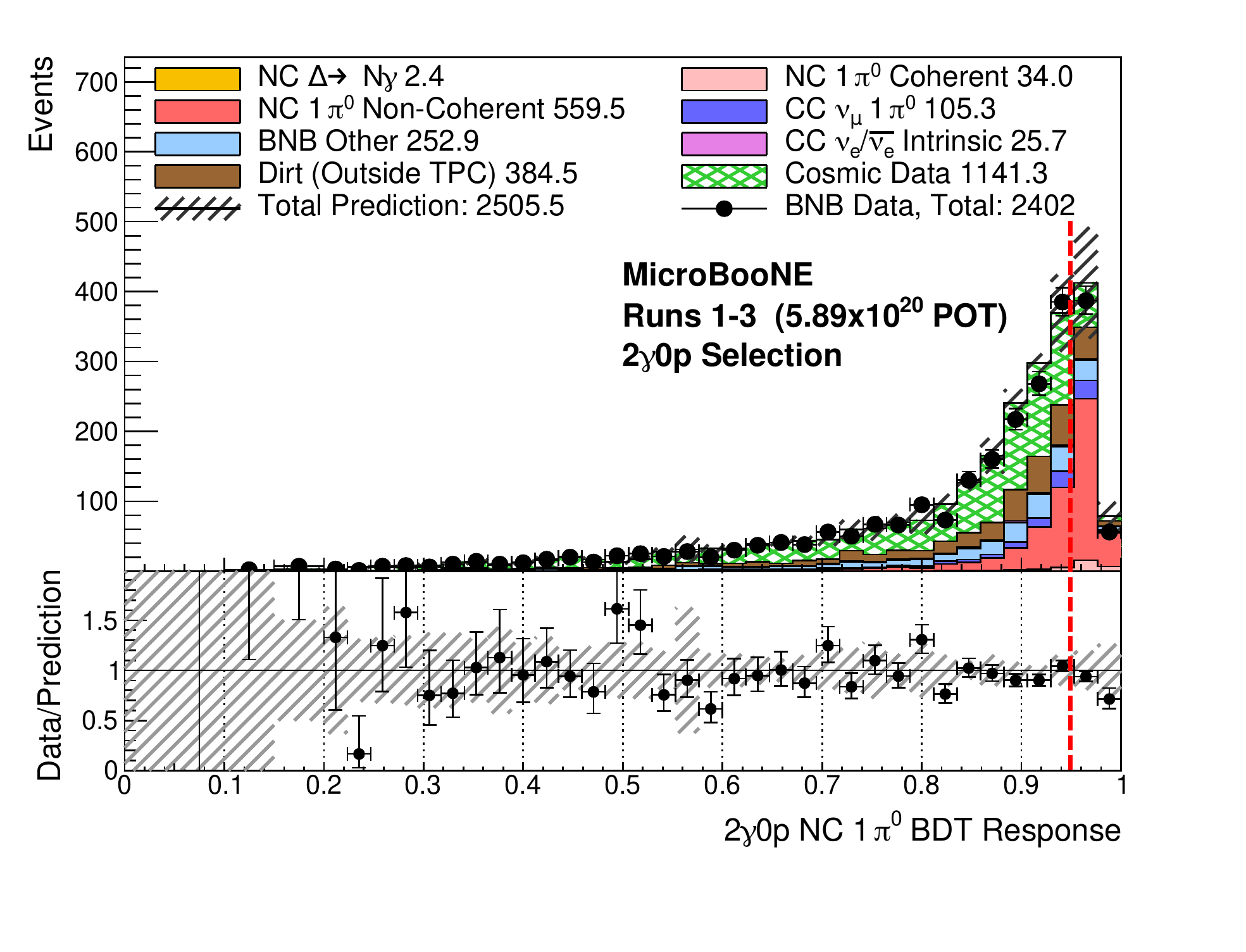}
      \vspace{-1cm}
        \caption{$2\gamma 0p$}
  \end{subfigure}
  \caption{The BDT classifier score for (a) $2\gamma 1p$ and (b) $2\gamma 0p$ targeted selections. Higher scores indicate more NC $1\pi^0$ signal-like events, and lower scores indicate more background-like events. The red vertical lines indicate the threshold positions, keeping all events to the right, for the final selections.}
  \label{fig:bdt_responses}
\end{figure}

By construction, BDT scores lie on the interval of [0, 1]. After training, the resulting BDT score distributions, tested on a statistically independent simulation and data set, are shown in Fig.~\ref{fig:bdt_responses}. The simulation and data points agree across the full range of BDT classifier score within systematic and statistical uncertainties (the definition of these systematic uncertainties is described in detail in Sec.~\ref{sec:systematics}). The bimodal distribution of the $2\gamma1p$ BDT response indicates greater separation power between signal and background compared to that for $2\gamma0p$ because the addition of the reconstructed track gives access to an entirely separate handle on background rejection. For this and subsequent MC simulation comparisons to data, the simulation predictions are broken down into the following eight categories, based on \textsc{genie} truth-level information:
\begin{itemize}
    \item NC 1$\pi^0$: All neutral current interactions that produce one exiting $\pi^0$ regardless of incoming neutrino flavor. This is our targeted signal selection, and it is further split into two sub-categories, ``NC 1$\pi^0$ Coherent'' and ``NC 1$\pi^0$ Non-Coherent'' contributions, based on their interaction types. Non-Coherent scattering occurs when a neutrino interacts with a nucleon inside the argon nucleus, potentially knocking out one or more nucleons. In coherent scattering the neutrino interacts with the nucleus as a whole, leaving it in its ground state. This interactions occurs with low momentum-transfer, and as such the resulting $\pi^0$ tends to be very forward relative to the incoming neutrino beam.
    \item NC $\Delta \rightarrow N\gamma$: Leading Standard Model source of NC single-photon production below 1~GeV originating from radiative decay of the $\Delta(1232)$ baryon.
    \item CC $\nu_\mu 1 \pi^0$: All $\nu_\mu$ CC interactions that have one true exiting $\pi^0$.
    \item CC $\nu_e/\overline{\nu}_{e}$ Intrinsic: All CC $\nu_e$ or $\overline{\nu}_{e}$ interactions regardless of whether or not a $\pi^0$ was emitted. 
    \item BNB Other: All remaining BNB neutrino interactions that take place in the active TPC volume of MicroBooNE and are not covered by the above five categories, such as multiple $\pi^0$ events and $\eta$ meson decay. See section \ref{sec:backgrounds} for more details.
    \item Dirt (Outside TPC): All BNB neutrino interactions that take place outside the MicroBooNE active TPC but have final states that enter and interact inside the active TPC detector. This can originate from scattering off liquid argon in the cryostat vessel outside the active TPC volume or from interactions in the concrete and ``dirt'' surrounding the cryostat itself. 
    \item Cosmic Data: Coincident cosmic ray interactions that take place during a BNB spill but without any true neutrino interaction present.
\end{itemize}

The final NC $1\pi^0$-enriched samples are selected by placing a requirement on the BDT score distribution that maximizes the product of NC $1\pi^0$ signal efficiency and purity. This corresponds to a threshold on the BDT scores of $>0.854$ and $>0.950$ for $2\gamma1p$ and $2\gamma0p$, respectively.  The final distributions are provided and discussed in Sec.~\ref{finaldistr}.

\subsection{Systematic Uncertainty Evaluation}
\label{sec:systematics}
Systematic uncertainties on the MC simulation prediction include contributions from uncertainties in the neutrino flux, the cross-section modeling, hadron re-interactions, detector effects, and the effect of finite statistics used in the background predictions (both simulations and cosmic ray data).

The flux systematic uncertainties incorporate hadron production uncertainties where the Booster proton beam hits the beryllium target, uncertainties on pion and nucleon scattering in the target and surrounding aluminum magnetic focusing horn of the BNB, and mismodeling of the horn current. Following \cite{MicroBooNE:2019nio}, 
these are implemented by reweighting the flux prediction according to neutrino type, parentage, and energy, and studying the propagated effects on the final event distributions. 

The cross-section uncertainties incorporate modeling uncertainties on the \textsc{genie} prediction~\cite{MicroBooNE:2021ccs,Andreopoulos:2009rq,GENIE:2021zuu}, evaluated by \textsc{genie} reweighting tools. The default \textsc{genie} uncertainties on NC resonant production arising from NC resonant vector and axial mass parameters of $m_V = 0.840 \pm 0.084 $ GeV and $m_A = 1.120 \pm 0.224 $ GeV, respectively, were assumed.  \textsc{genie} uses an effective cascade empirical model for hadronic final-state interactions, called \emph{hA2018}, which allows for reweighting to estimate the effect on final distributions. For more information on cross-section uncertainties in MicroBooNE, please see \cite{MicroBooNE:2021ccs}.

The hadron-argon reinteraction uncertainties are associated with the propagation of hadrons through the detector, as modeled in \textsc{geant4} \cite{GEANT4:2002zbu}. Both charged pions and proton reinteractions during propagation were considered and their impact estimated using the \textsc{geant4reweight} tool \cite{Calcutt:2021zck}.

The detector modeling and response uncertainties are evaluated using MicroBooNE's novel data-driven technique for assessing and propagating LArTPC detector-related systematic uncertainties \cite{MicroBooNE:2021roa}. This approach uses \textit{in situ} measurements of distortions in the TPC wire readout waveform signals -- caused by detector effects such as electron diffusion, electron drift lifetime, electric field, and the electronics response -- to parameterize these effects at the TPC wire level. This provides a detector model-agnostic way to study and evaluate their effects on the high level variables and, subsequently, the final event distributions. Additional detector systematics corresponding to variations in the charge recombination model, the scintillation light yield, and space charge effects~\cite{MicroBooNE:2019koz,MicroBooNE:2020kca} are separately evaluated and also included.

\subsection{Shower Energy Calibration} 

\begin{figure}
    \includegraphics[width=0.48\textwidth]{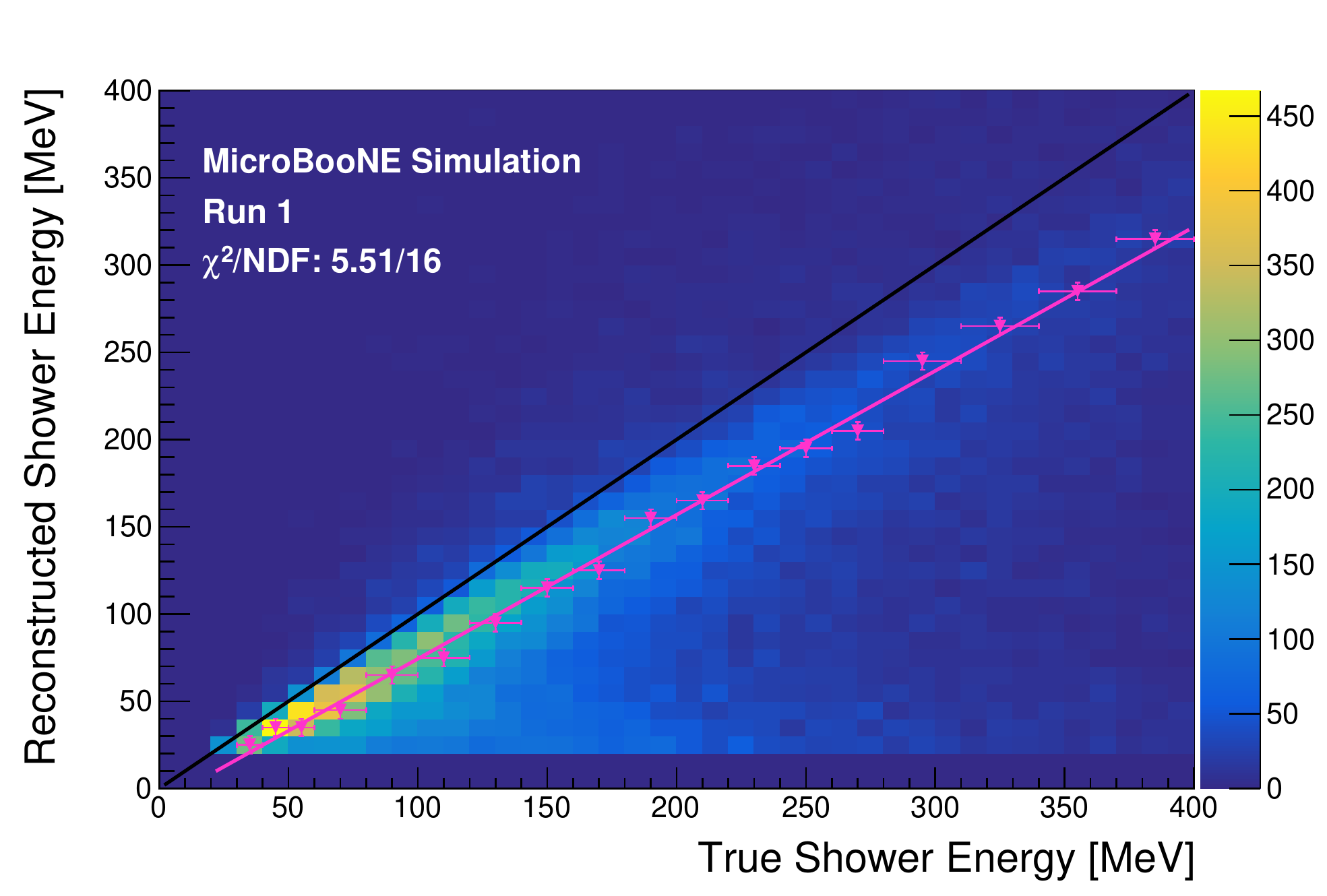}
    \caption{Reconstructed shower energy vs. true shower energy for a dedicated high statistics sample of simulated true NC $1\pi^0$ events. Only showers with a reconstructed energy of at least $20$~MeV are considered.}
    \label{fig:shower_energy_correction}
\end{figure}

Electromagnetic shower reconstruction in LArTPCs is known to be a lossy process primarily due to mis-clustering and thresholding effects. Current reconstruction algorithms often miss small, low-energy hits in an electromagnetic shower when clustering objects, and some of the hits that are reconstructed may fall below the energy threshold. On average, these effects are expected to yield shower energy losses of approximately 20\% \cite{caratelli2018}. This can be seen in Fig.~\ref{fig:shower_energy_correction}, which shows true and reconstructed shower energy for a dedicated high statistics sample of simulated true NC 1$\pi^0$ events, where the reconstructed shower energy falls systematically below the true shower energy in simulation. By performing a linear fit to the most probable values of reconstructed shower energy in bins of true shower energy, shown as the pink straight line in Fig.~\ref{fig:shower_energy_correction}, a correction factor is extracted which brings the reconstructed values closer to expectation. This fit results in an energy correction that is applied to all reconstructed showers,
\begin{equation}\label{eqn:ncpi0_energy_corr_factor}
    E_{\textrm{corr}} = (1.21 \pm 0.03)E_{\textrm{reco}} + (9.88 \pm 4.86) \ \textrm{MeV},
\end{equation}
and represents a correction of approximately 20\%, as expected.

\subsection{Final Selected Spectra}
\label{finaldistr}

\begin{figure}[h!]
    \centering 
    \begin{subfigure}{0.49\textwidth}
        \includegraphics[width = \textwidth]{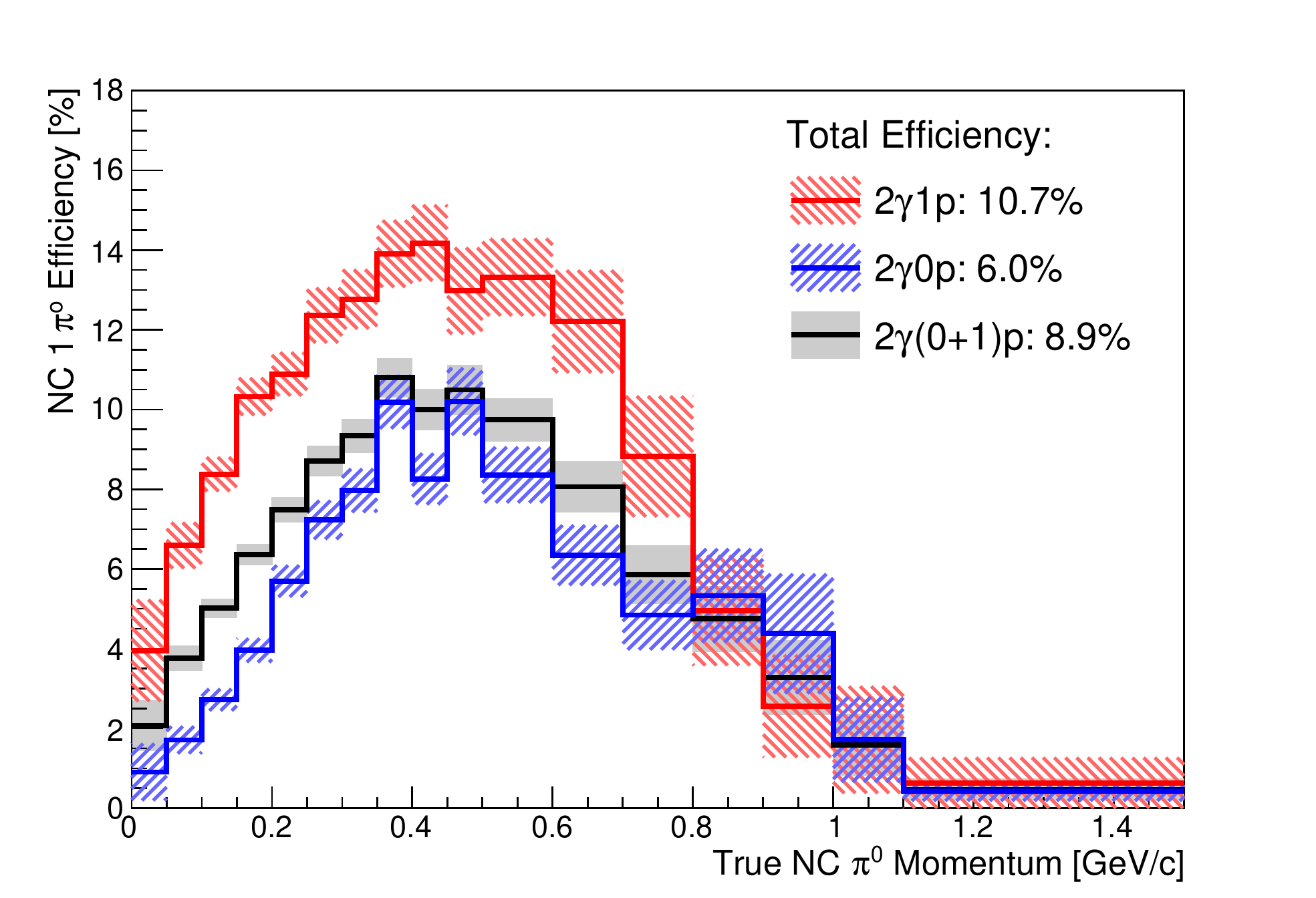}
        \label{fig:ncpi0_eff_mom}
        \caption{$\pi^0$ momentum dependence}
  \end{subfigure} 
  
  \begin{subfigure}{0.49\textwidth}
        \includegraphics[width = \textwidth]{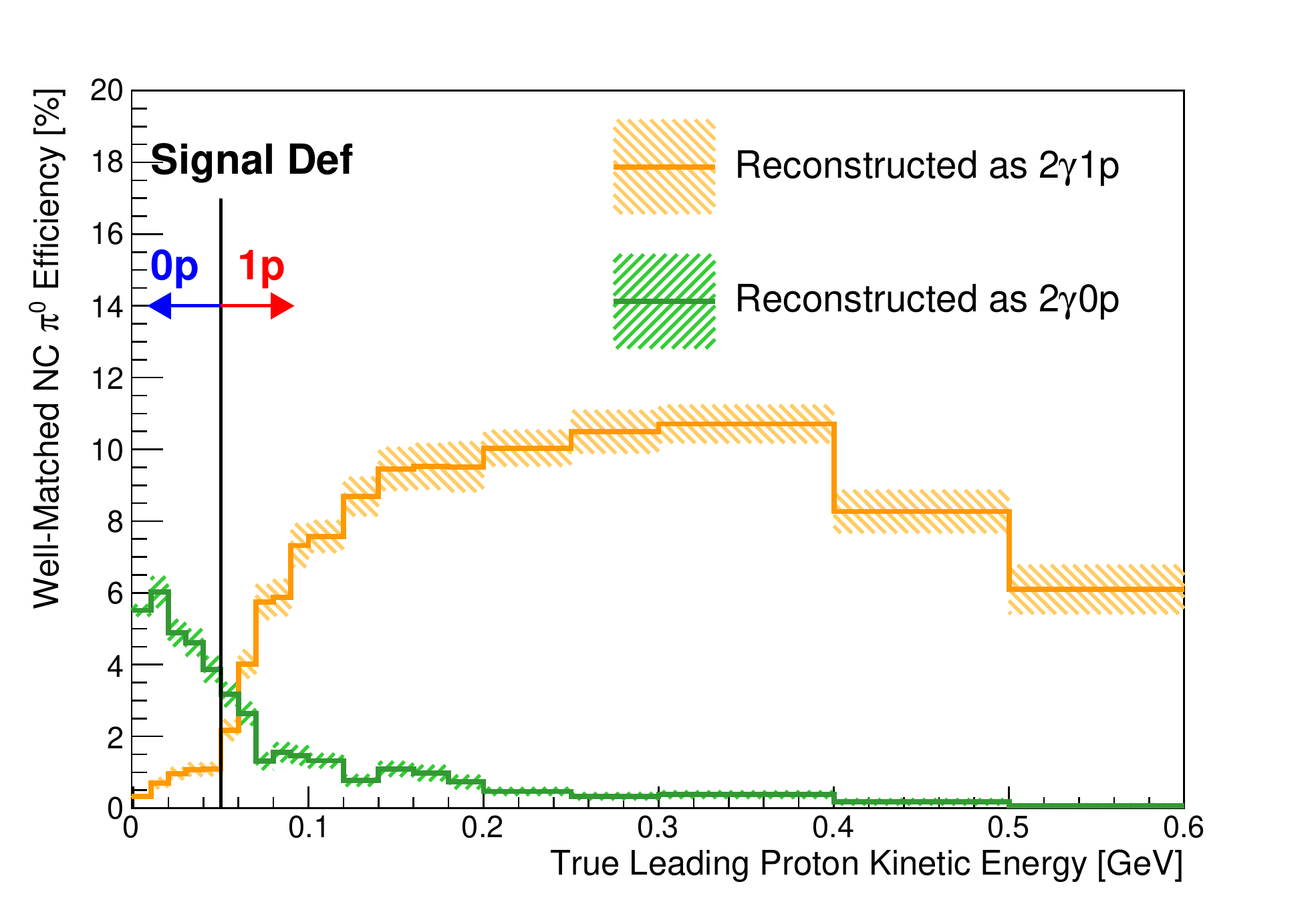}
        \label{fig:ncpi0_eff_pro}
        \caption{Proton kinetic energy dependence}
  \end{subfigure}
  \caption{(a) Efficiencies of the final $2\gamma1p$, $2\gamma0p$ and combined $2\gamma(0+1)p$ selections as a function of true $\pi^0$ momentum. (b) Efficiencies as a function true leading exiting proton kinetic energy for all NC1$\pi^0$ events that are reconstructed as either $1p$ or $0p$. Events in which there are no exiting protons are included in the first bin. As can be seen, a threshold of $\approx$ 50 MeV proton kinetic energy is where events start to shift between the $2\gamma0p$ and $2\gamma1p$ selections which was subsequently chosen as the signal definition for $0p$ and $1p$ signal events.}
  \label{fig:ncpi0_eff}
\end{figure}

After applying the BDT requirements, 1130 selected data events remain with 634 and 496 falling into the $2\gamma1p$ and $2\gamma0p$ selections, respectively. For the $2\gamma 1p$ selection, the BDT score requirement efficiency is 85.6\% and the purity is 63.5\% while for the $2\gamma 0p$ selection, the efficiency and purity are 58.8\% and 52.9\%, respectively. The $2\gamma 1p$ and $2\gamma 0p$ BDT selection efficiencies and purities are both calculated relative to their signal definition, with proton multiplicity counted with a kinetic energy threshold of 50 MeV. The efficiencies at each stage of the analysis are provided in Table~\ref{tab:efficiencies}, and the total efficiency for each selection is shown as a function of (a) true $\pi^0$ momentum and (b) true proton kinetic energy in Fig.~\ref{fig:ncpi0_eff}. Overall, the $1p$ selection is more efficient and of higher signal purity relative to the $0p$ selection due to the existence of a reconstructed particle track which greatly helps to tag the neutrino interaction point and reject backgrounds. This track information, particularly track calorimetry, provides an additional handle on the neutrino interaction mode; a proton-like track is highly indicative of an NC $1\pi^0$ interaction whereas CC interactions generally have a muon track in the final state.

\begin{figure}[h]
    \centering 
    \begin{subfigure}{0.49\textwidth}
        \includegraphics[trim={0 1.5cm 0 0},clip,width = \textwidth]{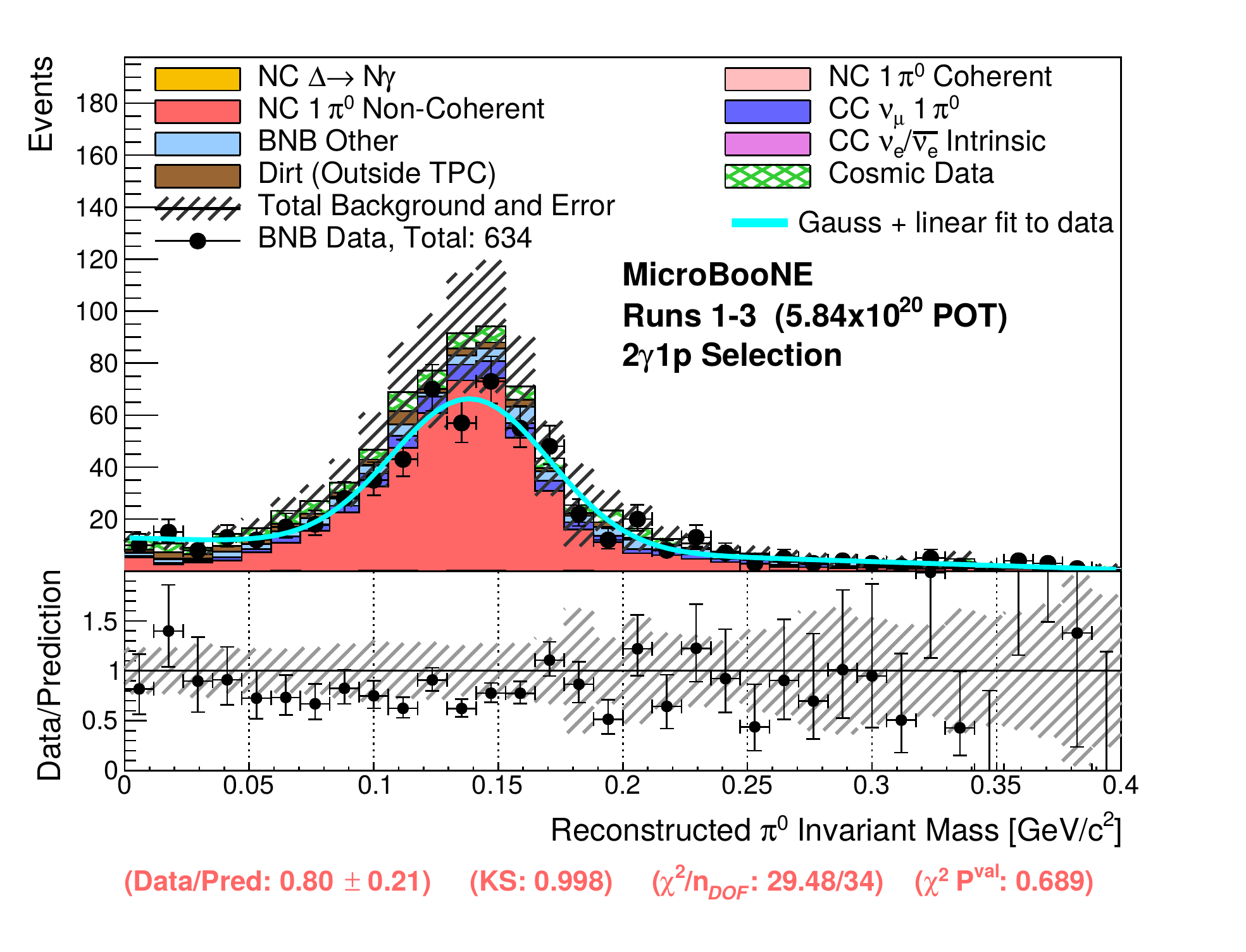}
        \caption{$2\gamma 1p$}
        \label{fig:ncpi0_invar_2g1p}
  \end{subfigure} 
  \begin{subfigure}{0.49\textwidth}
        \includegraphics[trim={0 1.5cm 0 0},clip, width = \textwidth]{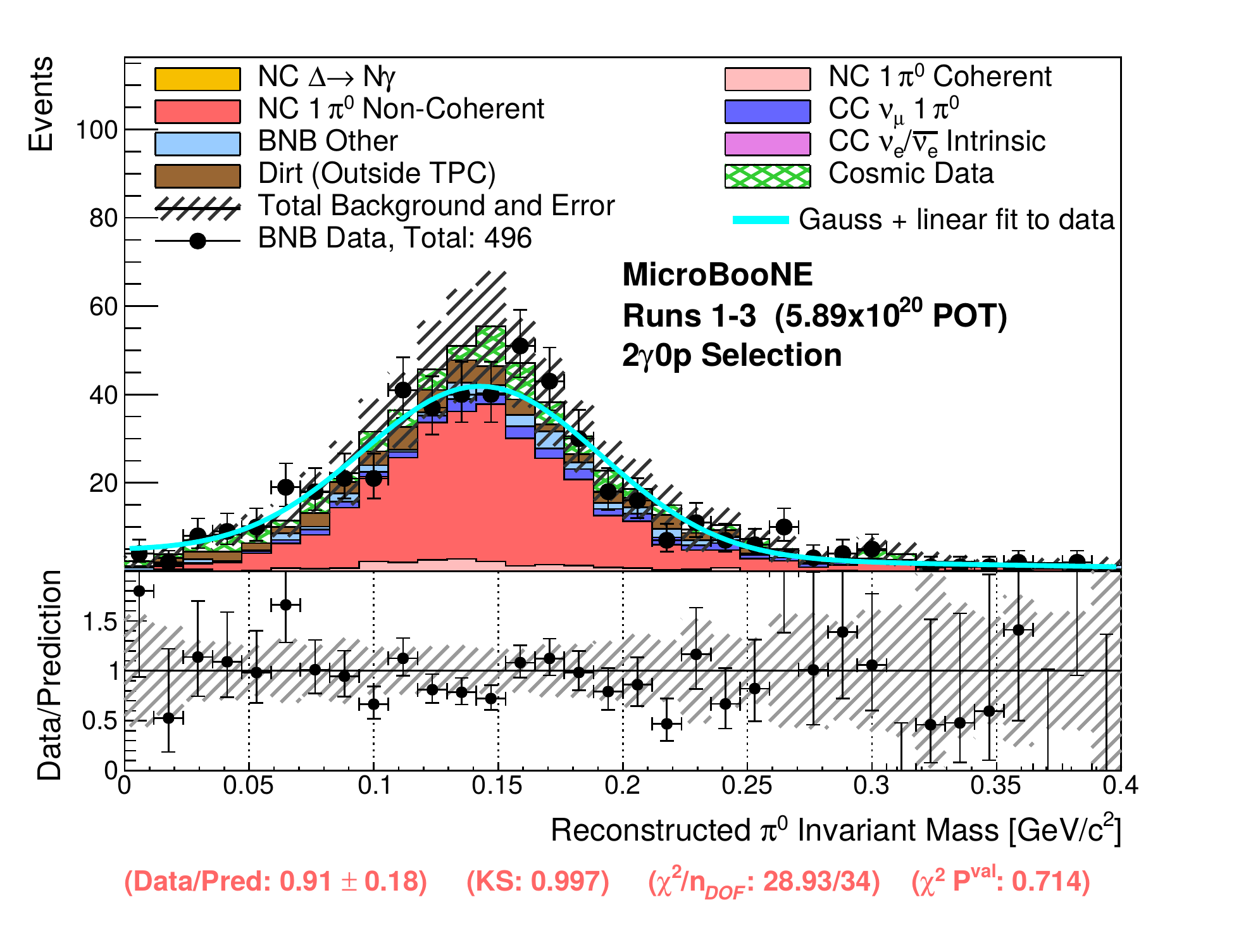}
        \caption{$2\gamma 0p$}
        \label{fig:ncpi0_invar_2g0p}
  \end{subfigure}
  \caption{The reconstructed diphoton invariant mass for both the (a) $2\gamma1p$ and (b) $2\gamma0p$ final selected data. The result of fitting a Gaussian plus linear function to the data is shown in cyan.}
  \label{fig:ncpi0_invar}
\end{figure}

\begin{table}[h!]
     \caption{NC $1\pi^0$ efficiencies for the $2\gamma1p$ and $2\gamma0p$ selections. The topological and combined efficiencies are evaluated relative to the defined exclusive $2\gamma1p$ and $2\gamma0p$ signal definitions, inside the active TPC. The pre-selection and BDT selection efficiencies are evaluated relative to their respective preceding selection stage. The final efficiencies are the combined total efficiency for each selection.}
    \begin{tabular}{lrrr}
    \hline\hline
        Selection Stage &  $2\gamma1p$ eff. & $2\gamma0p$ eff. \\ \hline
        Topological & 62.5\% & 47.4\%  \\
        Pre-selection & 19.9\% & 21.5\% \\
        BDT Selection & 85.6\% & 58.8\%  \\ \hline
        Final Efficiencies & 10.7\% & 6.0\%  \\ \hline\hline
    \end{tabular}
    \label{tab:efficiencies}
\end{table}

The resulting distributions as a function of the reconstructed two-photon invariant mass are shown in Fig.~\ref{fig:ncpi0_invar}. The invariant mass is reconstructed from the energy and direction of the two photon candidate showers as
\begin{equation}
   M_{\gamma \gamma}^2 = 2 E_{\gamma 1} E_{\gamma 2} (1- \cos \theta_{\gamma\gamma}),  
   \label{eq:invariant_mass}
\end{equation}
where $\cos \theta_{\gamma\gamma}$ is the opening angle between the two showers. For the $2\gamma1p$ case where a track has been identified as a candidate proton, the directions of the showers and thus the opening angle between them are calculated by constructing the direction between the candidate neutrino interaction point and the shower start point. For the $2\gamma0p$ selection, however, no such candidate track exists. Instead, the shower direction and opening angle are entirely estimated from the geometric shape of the showers themselves. 

\begin{figure*}[h!t]
    \centering 
    \begin{subfigure}{0.49\textwidth}
        \includegraphics[width = \textwidth]{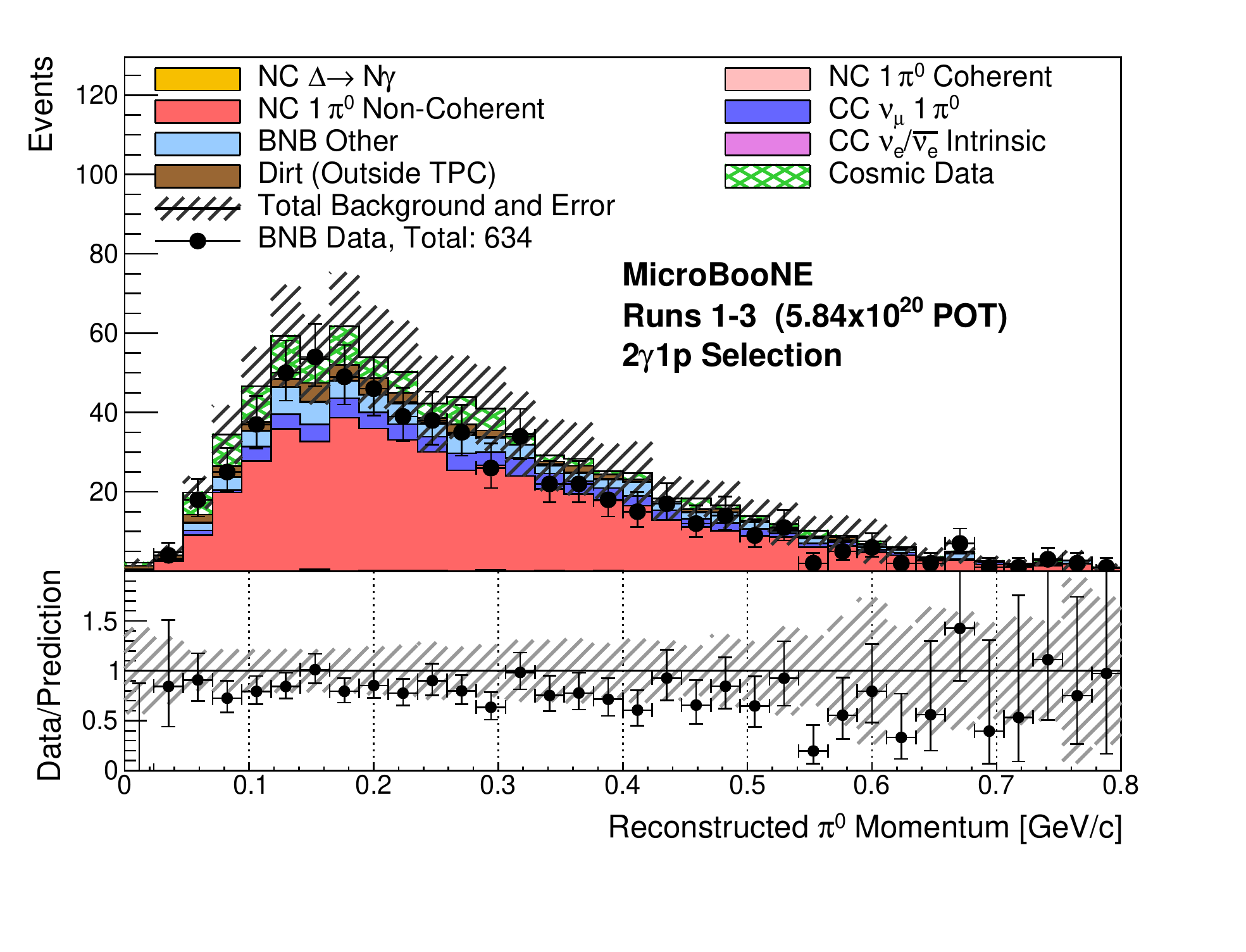}
        \vspace{-1cm}
        \caption{$2\gamma 1p$: Reconstructed $\pi^0$ momentum}
        \label{fig:ncpi0_mom_2g1p}
  \end{subfigure} 
    \begin{subfigure}{0.49\textwidth}
        \includegraphics[width = \textwidth]{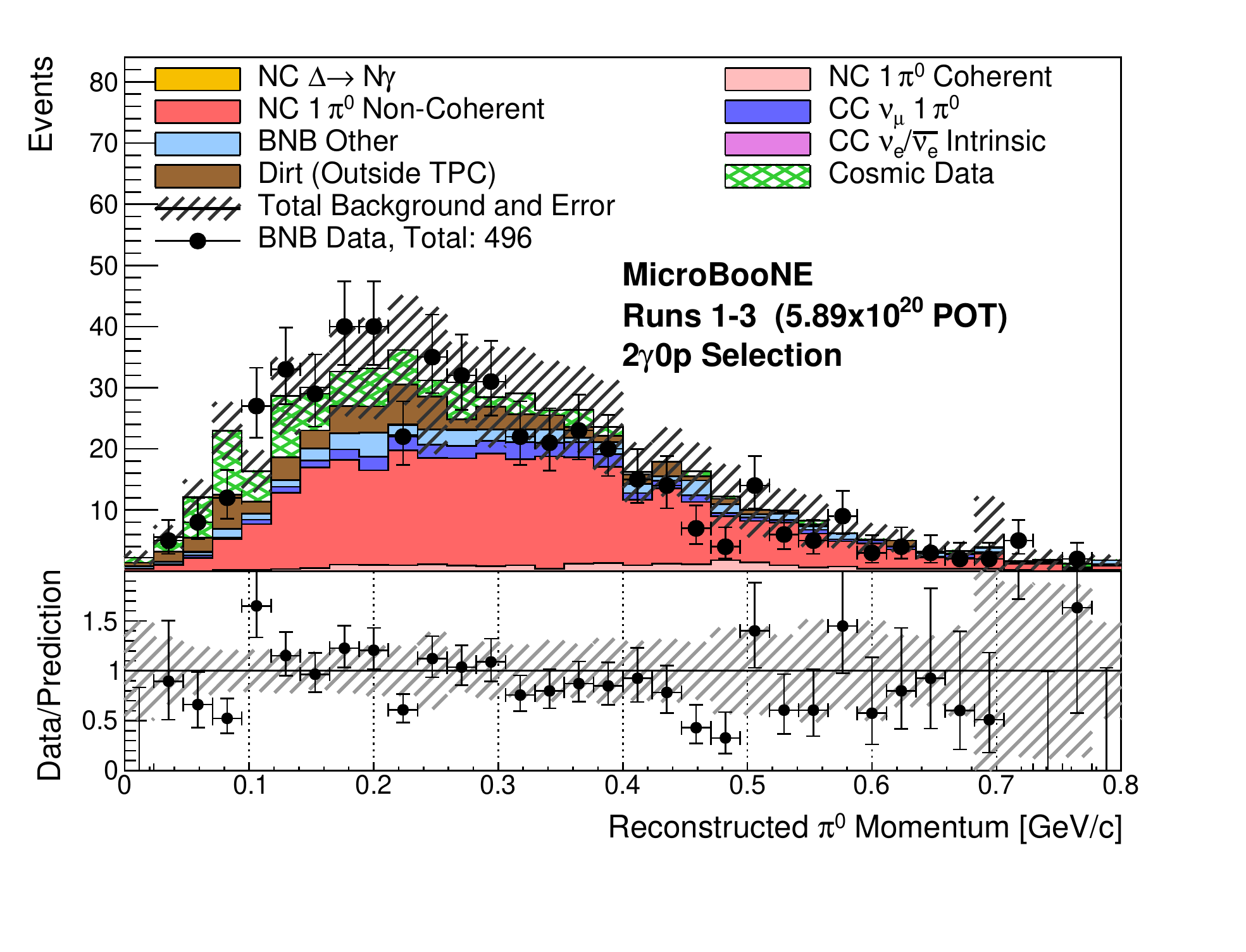}
        \vspace{-1cm}
        \caption{$2\gamma 0p$: Reconstructed $\pi^0$ momentum}
  \end{subfigure} 
    \begin{subfigure}{0.49\textwidth}
        \includegraphics[width = \textwidth]{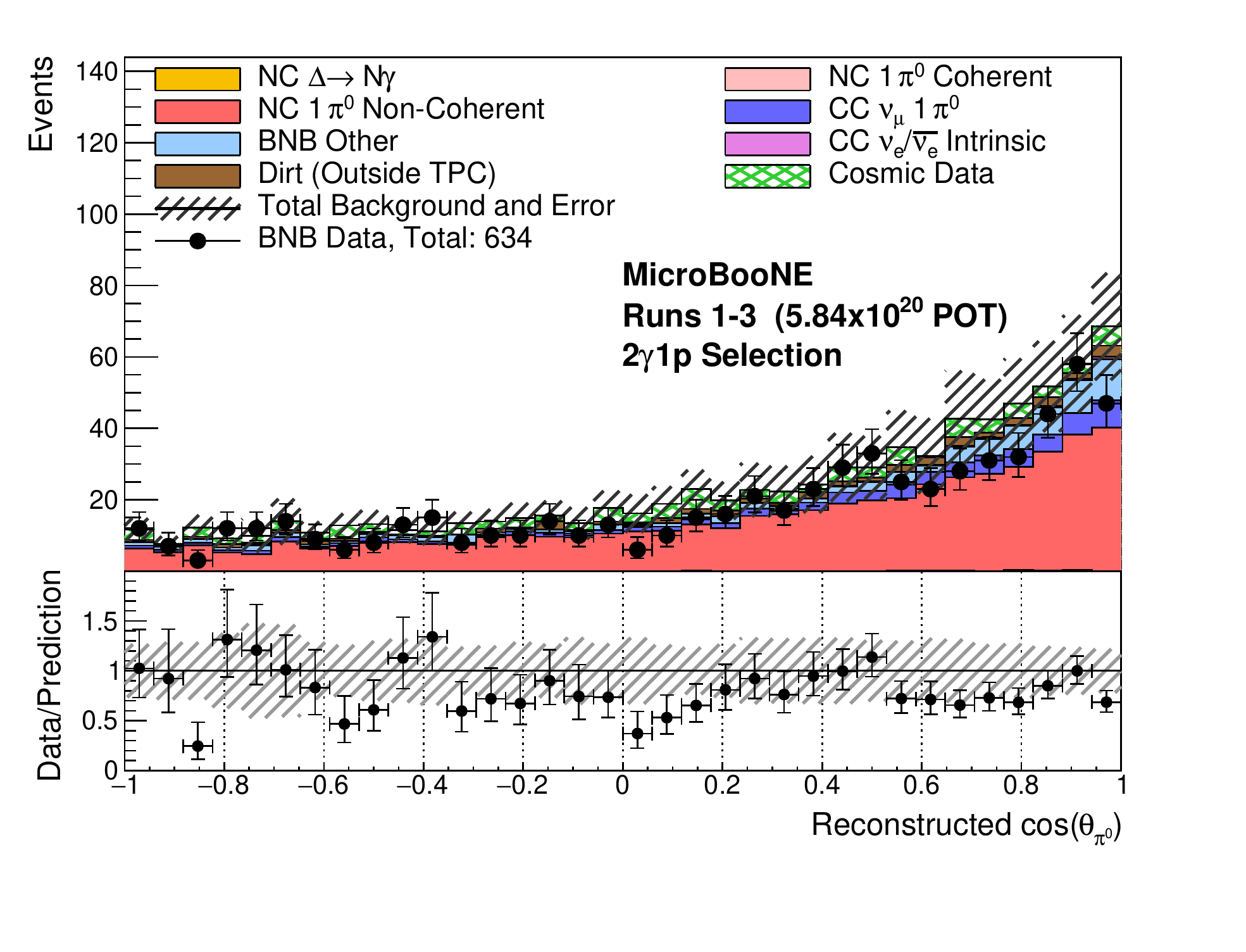}
        \vspace{-1cm}
        \caption{$2\gamma 1p$: Reconstructed cosine of $\pi^0$ angle}
  \end{subfigure} 
  \begin{subfigure}{0.49\textwidth}
        \includegraphics[width = \textwidth]{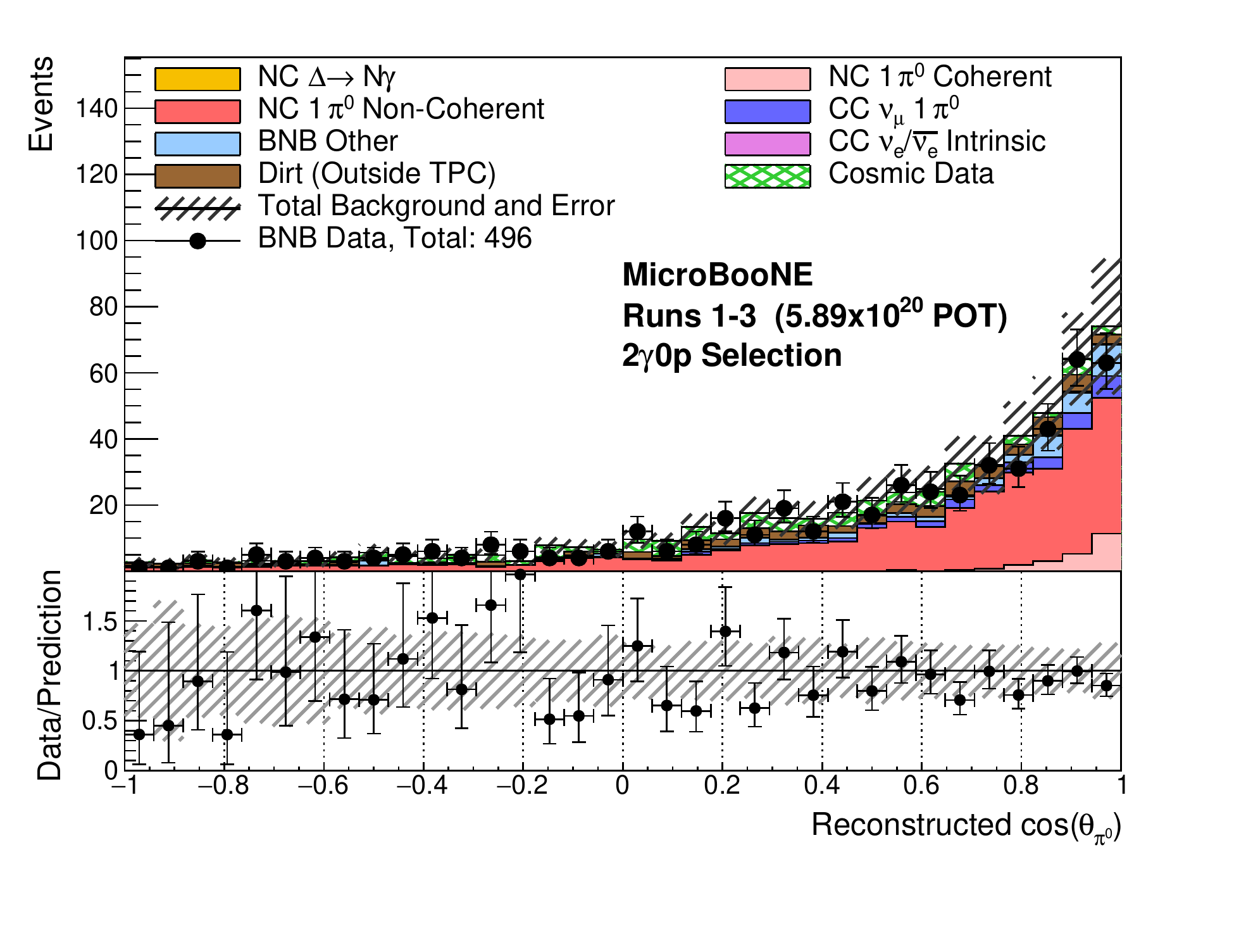}
        \vspace{-1cm}
        \caption{$2\gamma 0p$: Reconstructed cosine of $\pi^0$ angle}
  \end{subfigure}
  \caption{The reconstructed $\pi^0$ momentum ((a) and (b)) and reconstructed $\pi^0$ angle with respect to the neutrino beam ((c) and (d)) for both the $2\gamma1p$ ((a) and (c)) and $2\gamma0p$ ((b) and (d)) final selected data. The prediction shows agreement with the observed data within assigned uncertainties for the ranges shown, although a systematic deficit is observed in the total event rates as is discussed in Sec.~\ref{sec:deficit}.}
  \label{fig:ncpi0_costheta}
\end{figure*}

A Gaussian-plus-linear fit is performed to each observed distribution in data to extract the reconstructed $\pi^0$ invariant mass while taking into account the non-$\pi^0$ background contamination. For the $2\gamma 1p$ event sample, this fit gives a Gaussian mean of 138.9$\pm$2.1~MeV/$c^2$ with a width of 31.7$\pm$2.4~MeV/$c^2$. For the $2\gamma 0p$ event sample, the corresponding fit gives a Gaussian mean of 143.3$\pm$3.2~MeV/$c^2$ with a width of 47.9$\pm$4.9~MeV/$c^2$. As a goodness-of-fit test, the resulting $\chi^2$ per degree of freedom is 1.20 and 1.45 for the $2\gamma 1p$ and $2\gamma 0p$ fits, respectively. These both show agreement with the expected invariant mass of the $\pi^0$ of $134.9770\pm0.0005$ MeV/$c^2$ \cite{ParticleDataGroup:2020ssz} giving confidence and validation of the calorimetric energy reconstruction of the showers. Additional distributions showing the reconstructed $\pi^0$ momentum as well as the reconstructed angle of the outgoing $\pi^0$ with respect to the incoming neutrino beam are provided in Fig.~\ref{fig:ncpi0_costheta}. 

\begin{figure}[h]
    \centering 
    \begin{subfigure}{0.49\textwidth}
        \includegraphics[width = \textwidth]{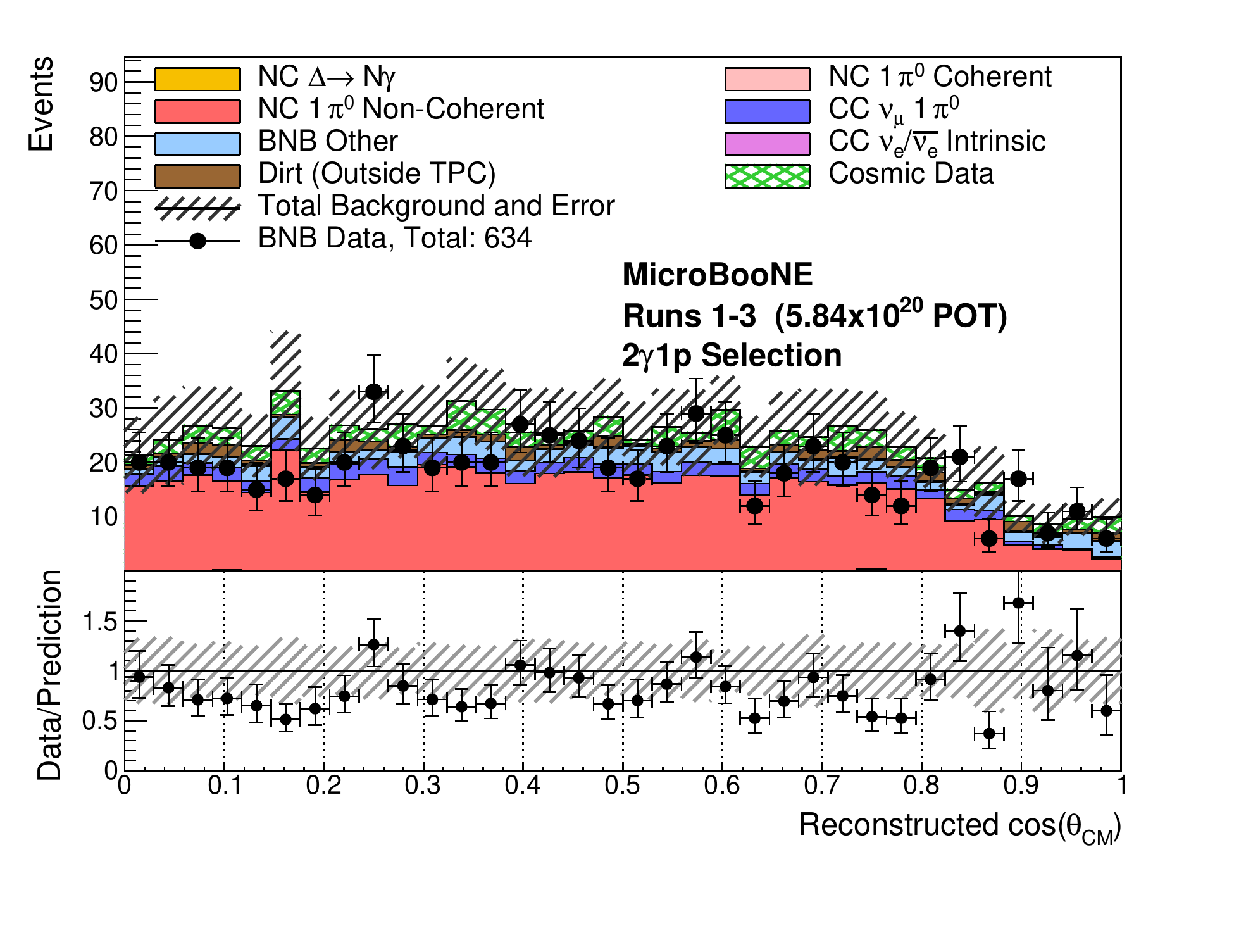}
        \vspace{-1cm}
       \caption{$2\gamma 1p$}
  \end{subfigure} 
  \begin{subfigure}{0.49\textwidth}
        \includegraphics[width = \textwidth]{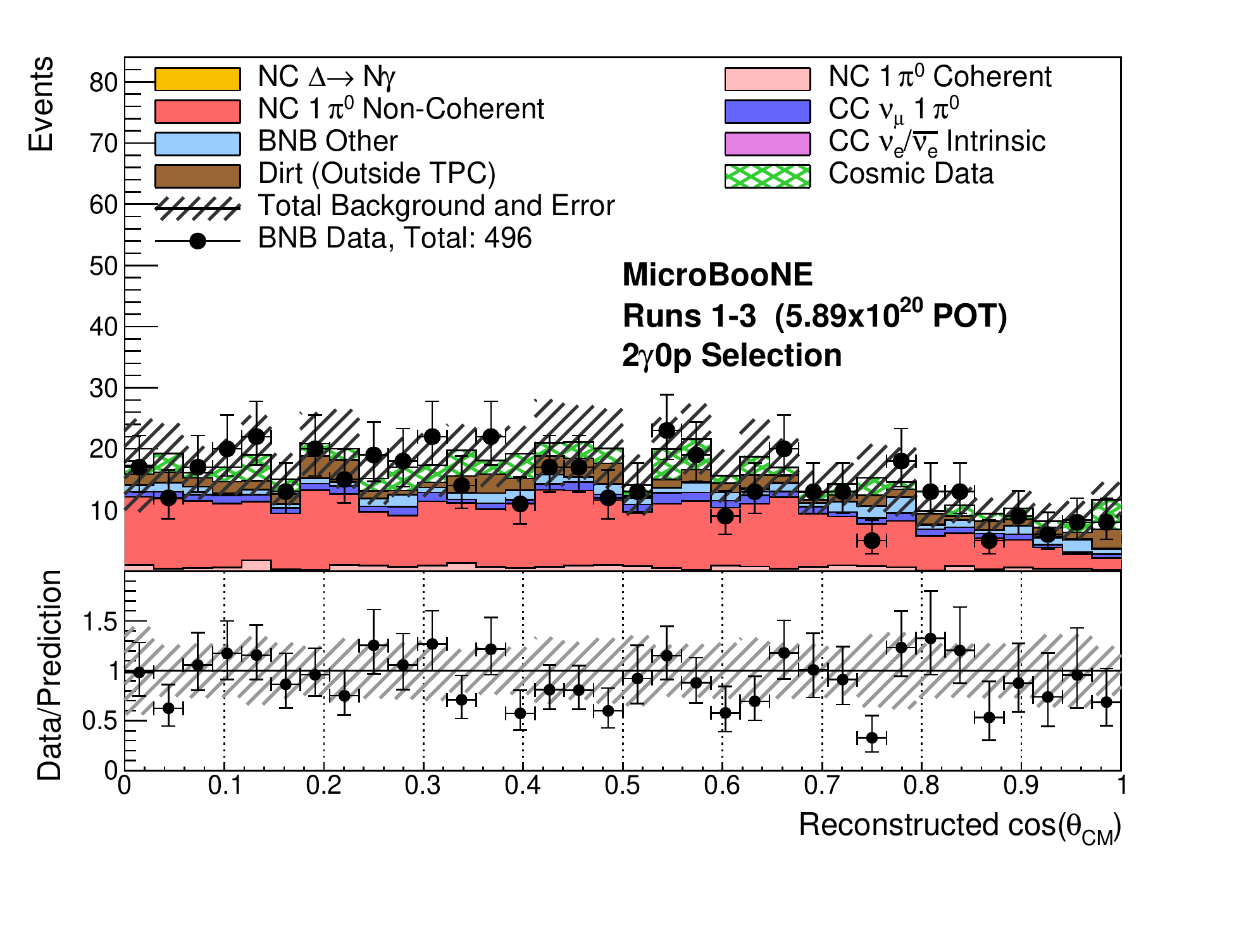}
        \vspace{-1cm}
        \caption{$2\gamma 0p$}
  \end{subfigure}
  \caption{The reconstructed cosine of the center-of-mass angle for the (a) $2\gamma1p$ and (b) $2\gamma0p$ final selected data.}
  \label{fig:ncpi0_costhetacm}
\end{figure}

Two additional reconstructed distributions of interest are highlighted. First, the reconstructed cosine of the center-of-mass (CM) decay angle is shown in Fig.~\ref{fig:ncpi0_costhetacm}. This is defined as the angle between the lab-frame $\pi^0$ momentum direction and the decay axis of the two daughter photons in the CM frame, 
\begin{equation}
\cos\theta_\text{CM} =\frac{|E_{\gamma1} - E_{\gamma2}|}{|p_{\pi^0}|}. 
\label{eq:cos_theta}
\end{equation}

\noindent This quantity should be an isotropic flat distribution for true $\pi^0 \rightarrow \gamma\gamma$ signal events, and any deviation from this can highlight regions of inefficiency in reconstruction or selection. 
As can be seen in Fig.~\ref{fig:ncpi0_costhetacm}, for both $2\gamma1p$ and $2\gamma0p$ selections, the distributions taper off at high $\cos \theta_\text{CM}$ which corresponds to increasingly asymmetric $\pi^0$ decays. When reconstructing asymmetric $\pi^0$ decay events, it is more likely that the subleading photon shower is missed due to its low energy. Note, however, that the observed data show the same trend as the simulation within uncertainty.  

\begin{figure}[h!]
    \centering 
    \begin{subfigure}{0.49\textwidth}
        \includegraphics[width = \textwidth]{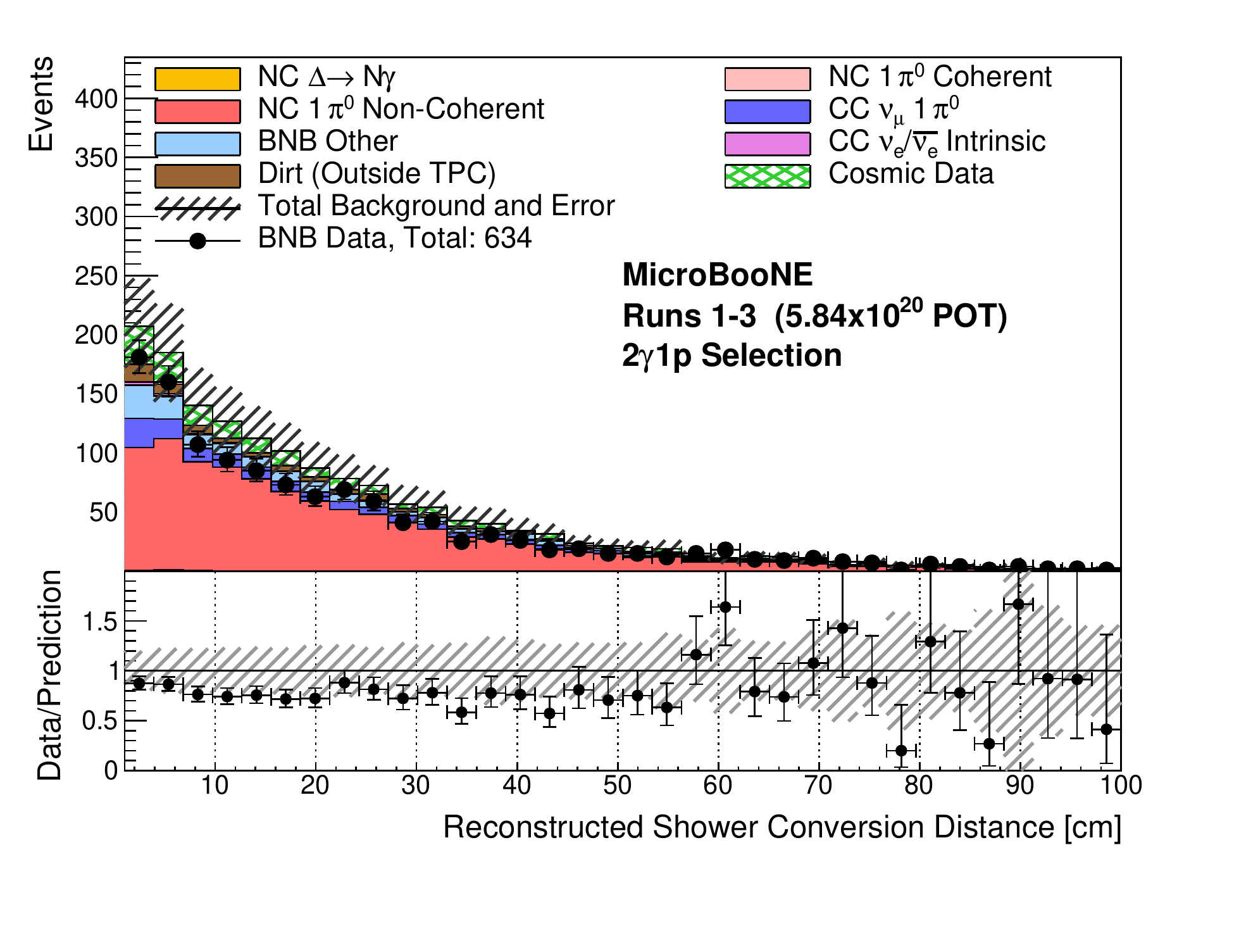}
  \end{subfigure} 
  \caption{The reconstructed conversion distance of both photons in the $2\gamma1p$ final selection.  There are well-reconstructed showers with conversion distances as far as 100~cm from the candidate neutrino interaction.}
  \label{fig:ncpi0_conv}
\end{figure}

Figure~\ref{fig:ncpi0_conv} additionally highlights the reconstructed photon conversion distance for all showers in the final $2\gamma1p$ selection.  Well-reconstructed showers with conversion distances as far as 100~cm from the candidate neutrino interaction are observed. This helps validate the assumption that the reconstructed showers are indeed likely to be true photons as $\mathcal{O}$(100)~MeV photons are expected to have a mean free path in argon of $\approx 20$~cm.  Note that, as the $2\gamma0p$ selection does not have any visible hadronic activity for tagging the interaction point, the corresponding conversion distance is significantly harder to estimate.

\begin{figure}
    \centering
    \begin{subfigure}{0.44\textwidth}
        \includegraphics[width=\textwidth]{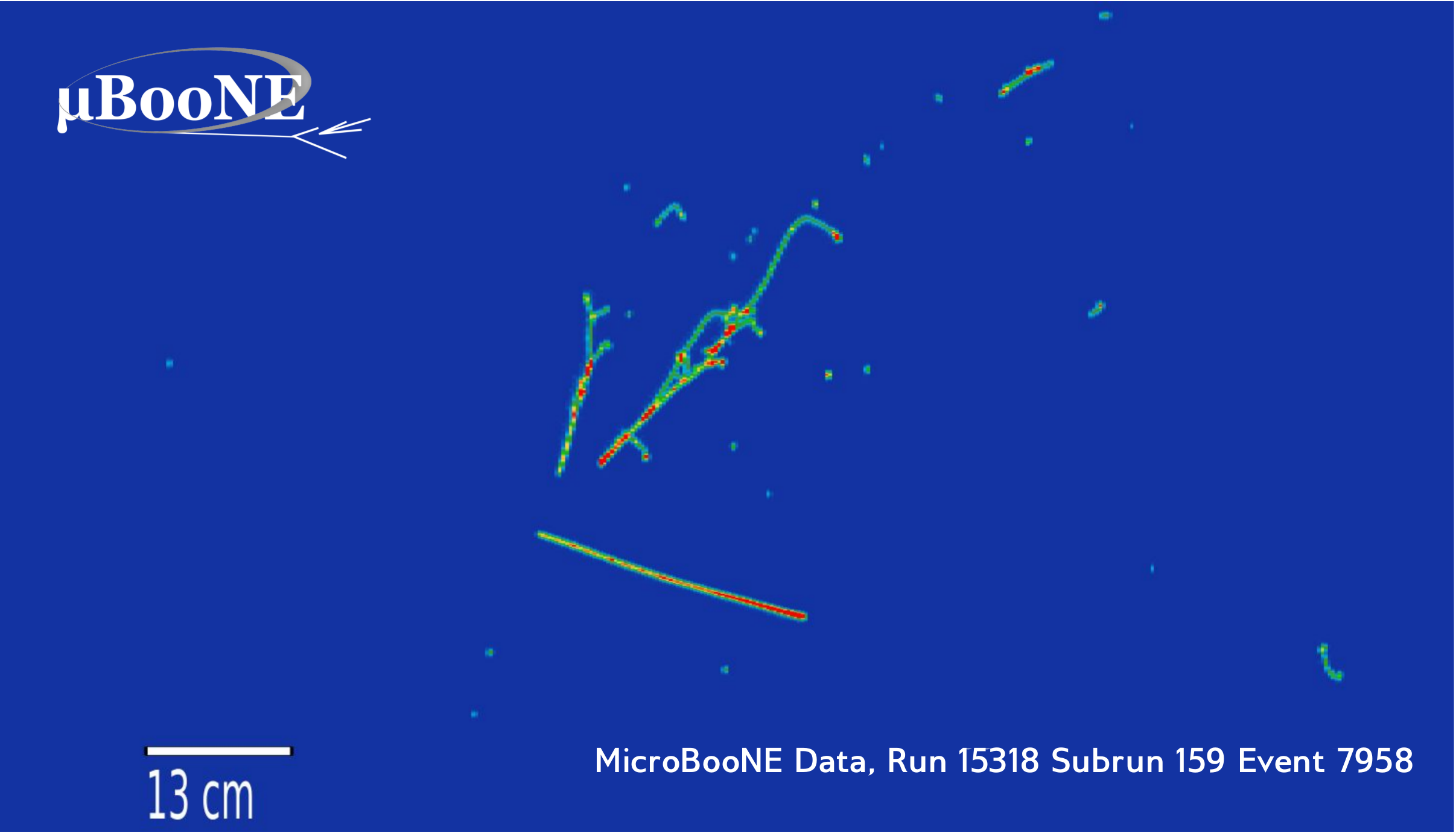}
        \caption{$2\gamma 1p$}
        \label{fig:2g1p_evd}
    \end{subfigure}
    \begin{subfigure}{0.44\textwidth}
        \includegraphics[width=\textwidth]{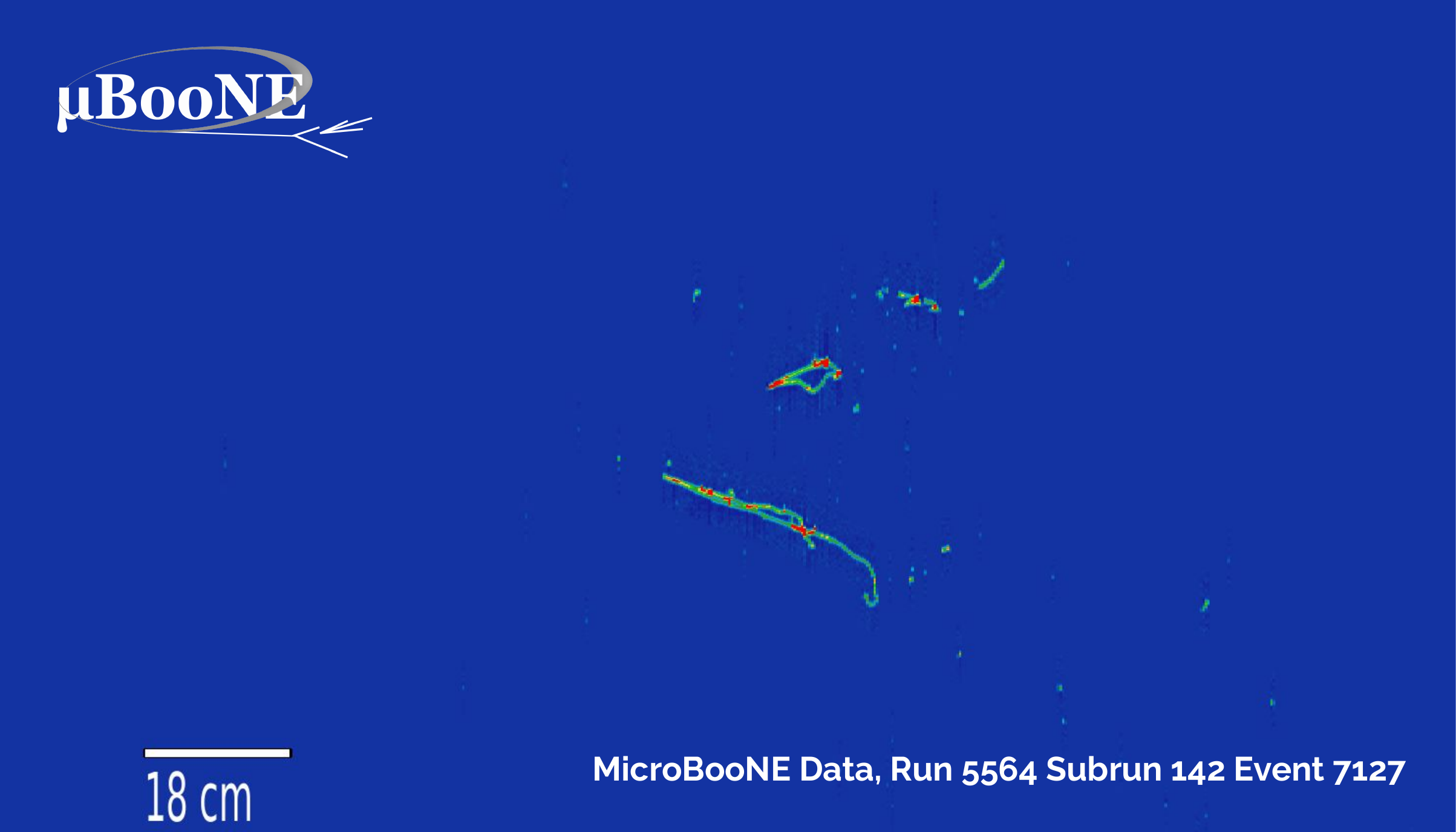}
        \caption{$2\gamma 0p$}
        \label{fig:2g0p_evd}
    \end{subfigure}
    \caption{Event displays of candidate NC $1\pi^0$ events found in the MicroBooNE data using (a) the $2\gamma 1p$ selection and (b) the $2\gamma 0p$ selection, on the MicroBooNE TPC collection plane. The horizontal axis here corresponds to the increasing wires, with an associated distance in cm. The vertical axis represents the TPC drift time. The aspect ratio of this plot is set such that the length scale shown for the horizontal axis is the same for the vertical axis.}
    \label{fig:ncpi0_evds}
\end{figure}

Finally, Fig.~\ref{fig:ncpi0_evds} shows two example event displays of selected events in data for both the $2\gamma 1p$ and $2\gamma 0p$ topologies. Each event shows two well-reconstructed showers pointing back to a common interaction point with properties consistent with those being photons from a $\pi^0$ decay.

\subsection{Background Discussion and Validation}
\label{sec:backgrounds}
\begin{figure}[h]
    \centering 
    \begin{subfigure}{0.49\textwidth}
        \includegraphics[trim={0 1.5cm 0 0},clip, width = \textwidth]{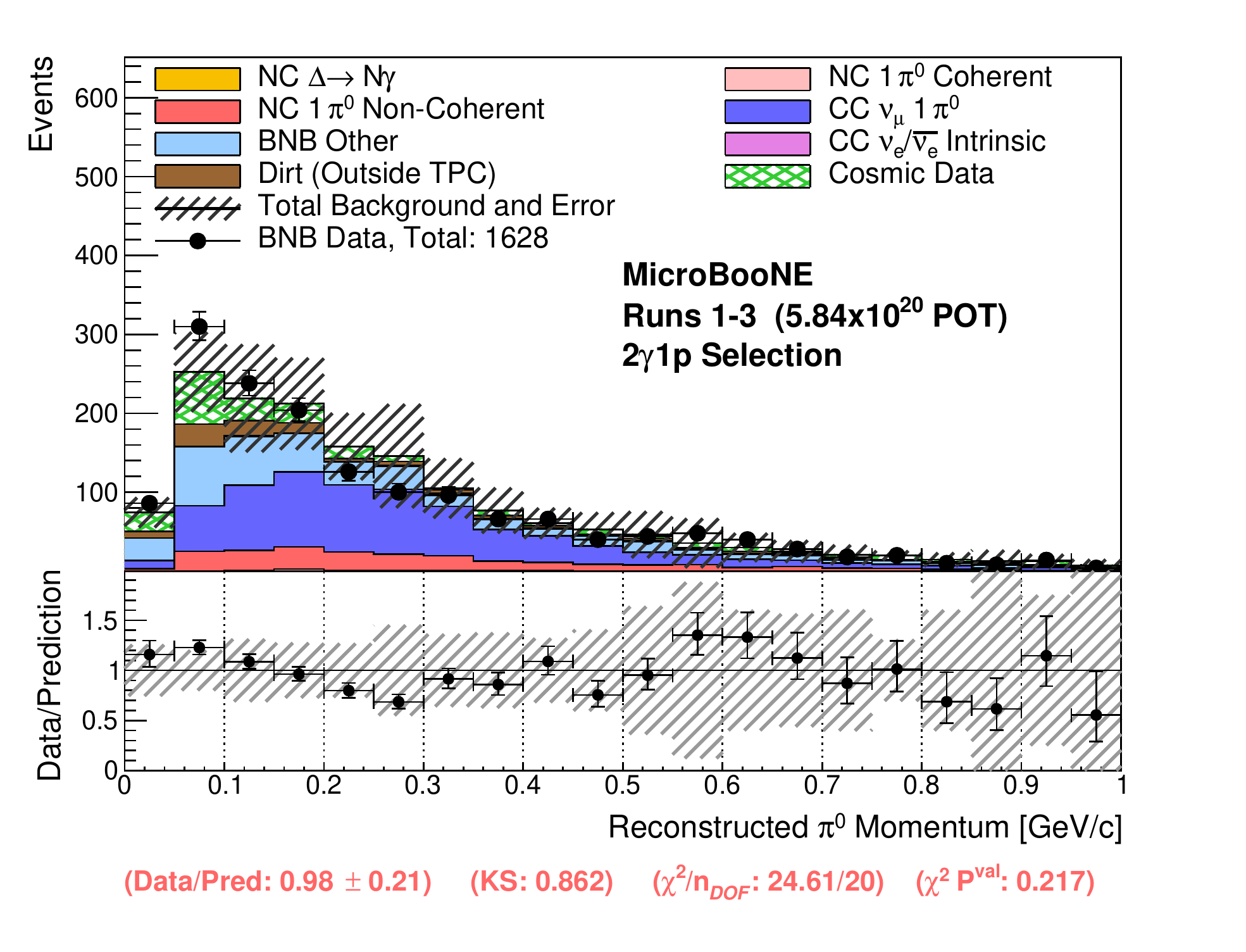}
       \caption{$2\gamma 1p$ background validation sample}
        \label{fig:bkg_study_1p}
  \end{subfigure} 
  \begin{subfigure}{0.49\textwidth}
      \includegraphics[trim={0 1.5cm 0 0},clip, width = \textwidth]{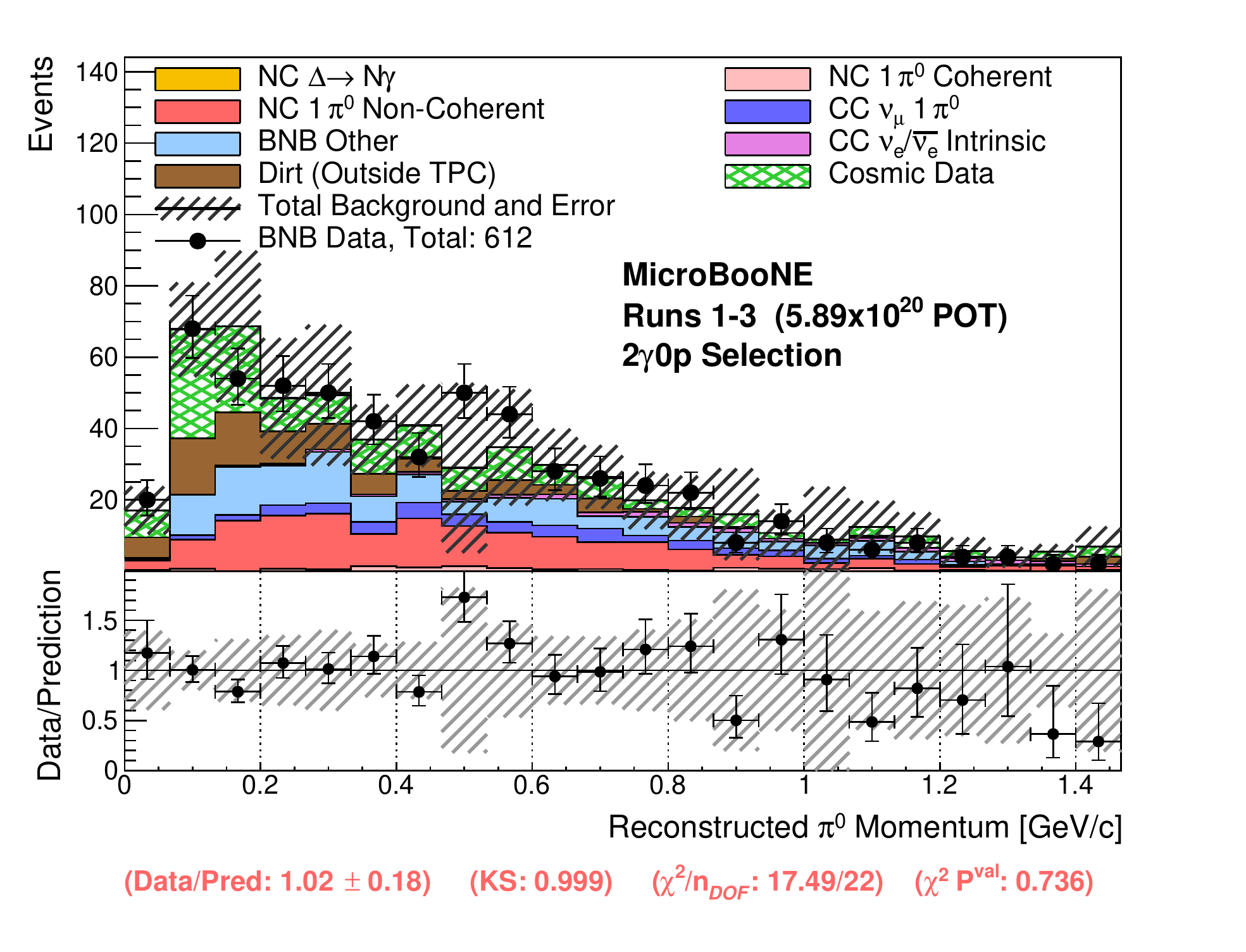}
        \caption{$2\gamma 0p$ background validation sample}
        \label{fig:bkg_study_0p}
  \end{subfigure}
  \caption{Background rich sidebands for validation created by inverting the BDT selection cuts, for (a) $2\gamma1p$ and (b) $2\gamma0p$ samples. We observe good agreement between the data and prediction, within the assigned uncertainties giving us confidence in the background modelling.}
  \label{fig:bkg_study}
\end{figure}

In order to validate the background modelling in this analysis we developed a background rich sideband selection by inverting the BDT score cuts shown in Fig.~\ref{fig:bdt_responses}. This gives us a high statistics sample of ``CC$1\pi^0$'' and ``BNB Other'' background categories with which to compare to data. In addition to inverting the BDT score an additional cosmic rejection cut, where we require the Pandora neutrino score to be $>0.5$, is applied to provide a higher purity of the backgrounds we wish to study. These distributions are shown in Fig.~\ref{fig:bkg_study_1p} and ~\ref{fig:bkg_study_0p} for $2\gamma1p$ and $2\gamma0p$ respectively. This is particularly useful for $2\gamma1p$ inverted selection as the resulting spectrum is rich in CC$1\pi^0$ for higher reconstructed $\pi^0$ momenta, and richer in ``BNB Other'' at low momenta. For the $2\gamma0p$ background rich sample there is still a significant amount of rejected NC $\pi^0$ signal events, but the enhanced backgrounds still provide additional validation. The data is observed to be in good agreement with the prediction, within assigned uncertainties, with a $\chi^2/\text{ndof}$ of $24.61/20$ and $17.49/22$ for $2\gamma1p$ and $2\gamma0p$ respectively. This gives us confidence that the backgrounds and uncertainties are sufficiently modelled for a cross-section extraction to proceed.

We can also break down the ``BNB Other'' category further in order to improve our understanding of this important background. We find that approximately $75\%$ of ``BNB Other'' events contain true photons reconstructed, and in case of $2\gamma1p$ 84\% have true protons reconstructed. This indicates that despite being a background category the BDT's are indeed selecting events with a very high purity of true photons and protons, as is the target of the selection. The approximate breakdown of ``BNB Other'' after applying the BDT cuts is found to be

\begin{itemize}
    \item $\approx 25\%$ are events in which multiple $\pi^0$ are exiting the nucleus but only 1 $\pi^0$ is reconstructed correctly,
    \item $\approx 25\%$ are events in which no $\pi^0$ exits the nucleus but due to baryon or charged pion re-scattering in the argon, a $\pi^0$ is subsequently created and reconstructed,  
    \item $\approx 25\%$ are events in which there is no $\pi^0$ and a NC proton is reconstructed as the track with cosmic contamination resulting in a $2\gamma$ event mimicking a $\pi^0$,
    \item $\approx 20\%$ are events containing a NC $\eta \rightarrow \gamma \gamma$ decay event. These are generally rejected due to them having higher energies, but a small number are selected as they are topologically identical to the $\pi^0\rightarrow \gamma\gamma$ decay signal, 
    \item $\approx 5\%$ Miscellaneous other reconstruction failures, representing less than 1\% of total background events.  
\end{itemize}

In the case of CC$1\pi^0$ we see a similar situation, with 97.3\% (98.1\%) of $2\gamma1p$  ($2\gamma0p$) background events having at least 1 shower matched to a true $\pi^0$ and with 78\% of tracks in the $2\gamma1p$ sample being correctly matched to a proton. As such, the vast majority of these events have the correct target particle content, but rather the muon itself is missed. This primarily occurs when the muon is incorrectly clustered into a photon electro-magnetic shower due to close proximity, or when the muon is correctly reconstructed but is mis-identified as a cosmic muon, with the associated $\pi^0$ then being reconstructed as an isolated neutrino event.

\section{NC $\pi^0$ Rate Validation }
\label{sec:deficit}

NC $1\pi^0$ events contribute as a dominant background to NC single-photon production measurements carried out or planned by MicroBooNE such as searches for NC $\Delta$ radiative decay \cite{MicroBooNE:2021zai}, NC coherent single-photon production, or more rare $e^{+}e^{-}$ pair production motivated in BSM theories. In addition to using these selected events as a calibration sample for understanding and validating shower reconstruction performance, they are also used to validate the observed overall rate of this process as currently modeled with \textsc{genie}. Assuming \textsc{genie} provides a sufficient description of the observed data, this sample can and has been used to provide an \textit{in situ} constraint on NC $1\pi^0$ mis-identified backgrounds, \textit{e.g.}~as in \cite{MicroBooNE:2021zai}. Alternatively, these measurements can be used to increase our understanding of our current NC $\pi^0$ modelling and potentially motivate \textsc{genie} tuning.

As shown in Fig.~\ref{fig:ncpi0_costheta}, both the $2\gamma1p$ and the $2\gamma0p$ selections see an overall deficit in data relative to the MC prediction. This is more pronounced in the $2\gamma1p$ selection where the ratio of the number of selected data events to the number of selected simulated events is 0.79. As it is also known that the \textsc{genie} branching fraction of coherent NC $1\pi^{0}$ production on argon is significantly lower than expectation extrapolated from MiniBooNE's $\pi^{0}$ measurement on mineral oil~\cite{MiniBooNE:2008mmr}, the possibility of a correction to \textsc{genie} predictions on both non-coherent and coherent NC $1\pi^0$ production is explicitly examined. The MC predictions are fitted to data allowing both coherent and non-coherent NC $1\pi^{0}$ rates to vary. Both normalization-only and normalization plus shape variations to the coherent and non-coherent rates are explored; all yield similar conclusions. This section describes the normalization plus shape variation fit in detail.  

The normalization plus shape variation fit is performed as a function of reconstructed $\pi^0$ momentum for both 2$\gamma1p$ and $2\gamma0p$ selections, using [0, 0.075, 0.15, 0.225, 0.3, 0.375, 0.45, 0.525, 0.6, 0.675, and 0.9] GeV/c bin limits. In the fit, MC predicted coherent NC $1\pi^0$ events are scaled by a normalization factor $N_{coh}$, and MC predicted non-coherent NC $1\pi^0$ events are scaled on an event-by-event basis depending on their corresponding true $\pi^0$ momentum according to $(a+b|\vec{p}^{~true}_{\pi^{0}}|)$, where the true $\pi^0$ momentum is given in [GeV/c]. This linear scaling as a function of $\pi^0$ momentum was chosen because it was the simplest implementation that was consistent with the observed data-to-MC deficit, as observed in Fig.~\ref{fig:ncpi0_mom_2g1p}.

At each set of fitting parameters ($N_{coh}$, $a$, $b$), a $\chi^2$ is evaluated between the scaled prediction for this parameter set and the observed data using the Combined-Neyman-Pearson $\chi^2$ \cite{Ji:2019yca}. The $\chi^2$ calculation makes use of a covariance matrix including statistical and systematic uncertainties and correlations corresponding to the scaled prediction. Flux, cross section, detector and \textsc{geant4} systematic uncertainties are included in the fit including bin-to-bin systematic correlations. As the goal of the fit is to extract the normalization and scaling parameters of the coherent and non-coherent NC $1\pi^{0}$ rates, the cross-section normalization uncertainties of coherent and non-coherent NC $1\pi^{0}$ are not included. Note that the cross-section normalization uncertainties of coherent and non-coherent NC $1\pi^{0}$ are only removed for the purposes of this fit and not for the cross section extraction described in the following section. 

\begin{figure}[h!]
  \begin{subfigure}{0.49\textwidth}
  \includegraphics[width = \textwidth]{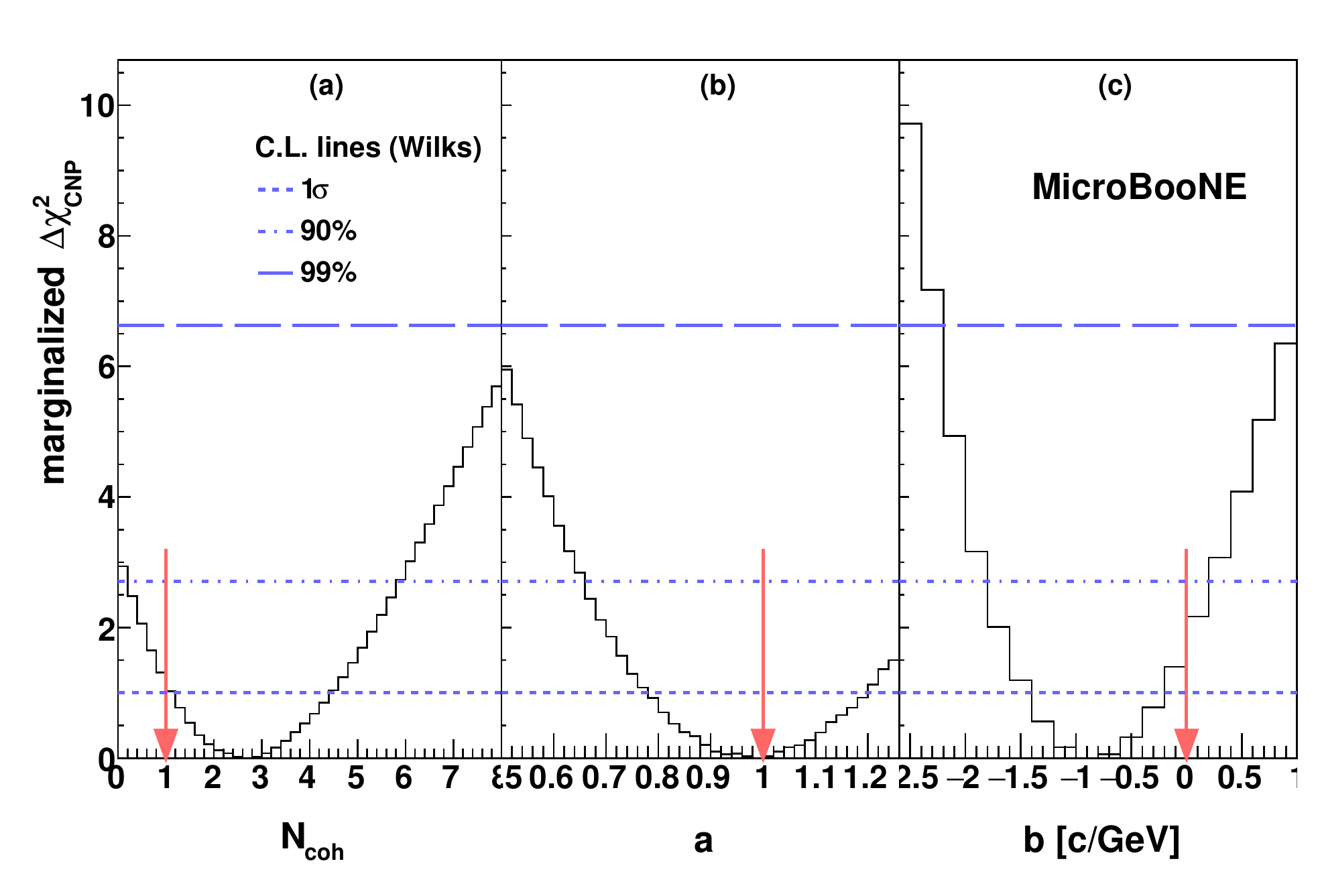}
  \end{subfigure}
\caption{The distribution of marginalized $\Delta\chi^2$, as a function of flat normalization factor for (a) coherent NC 1$\pi^0$ momentum-independent scaling factor, (b) non-coherent NC 1$\pi^0$ momentum-independent scaling factor, and (c) coefficient of momentum-dependent scaling factor for NC non-coherent 1$\pi^0$, marginalized over the other two parameters. The red arrows indicate parameter values expected for the \textsc{genie} central value prediction. The 1$\sigma$, 90\% and 99\% C.L.~lines are based on the assumption that the distribution follows a $\chi^2$ distribution with 1 degree of freedom.
}
\label{fig:marginalized_chi}
\end{figure}

\begin{figure}[h!]
  \begin{subfigure}{0.49\textwidth}
   \includegraphics[width = \textwidth]{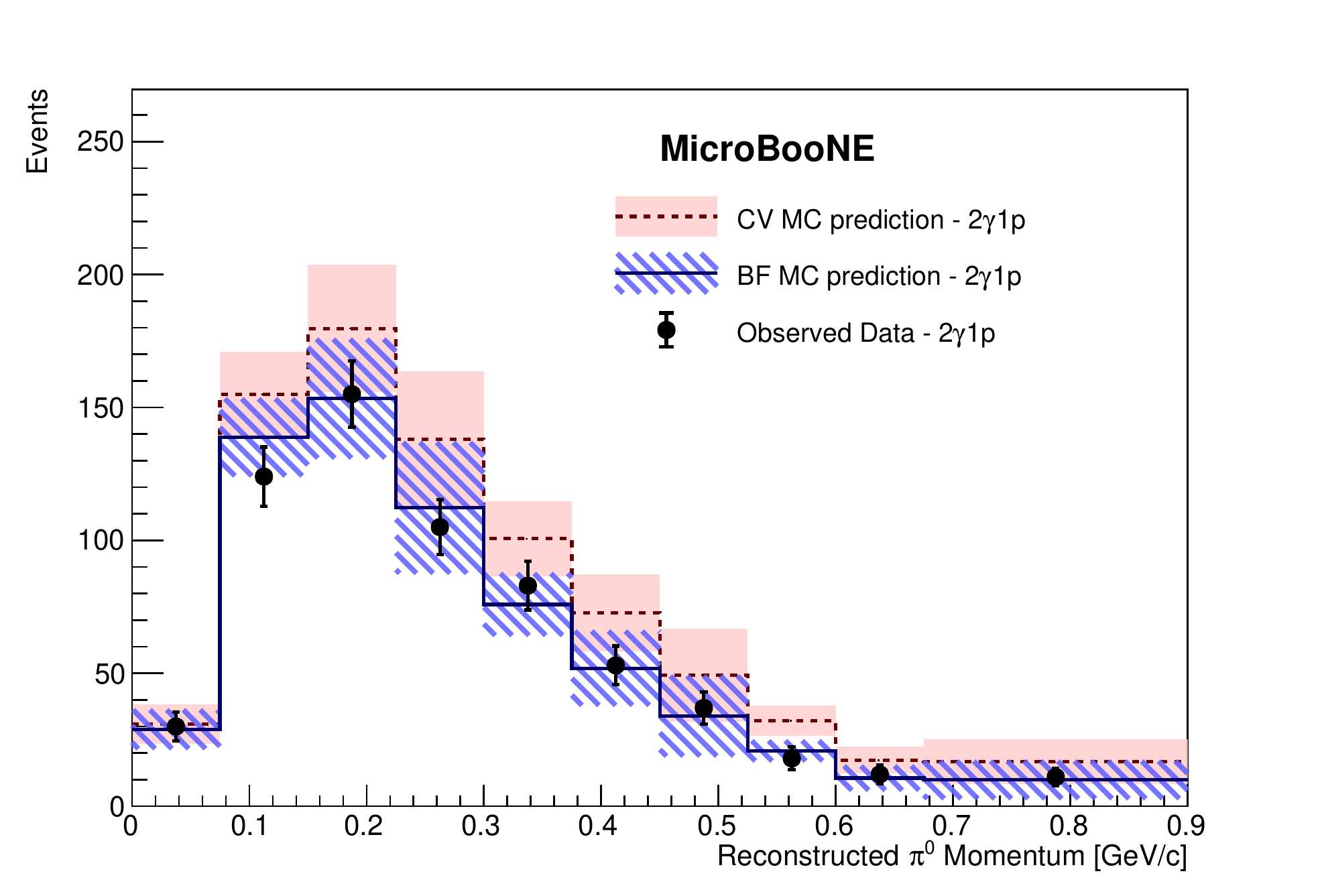}
  \caption{$2\gamma1p$}
  \end{subfigure}%
  
  \begin{subfigure}{0.49\textwidth}
   \includegraphics[width = \textwidth]{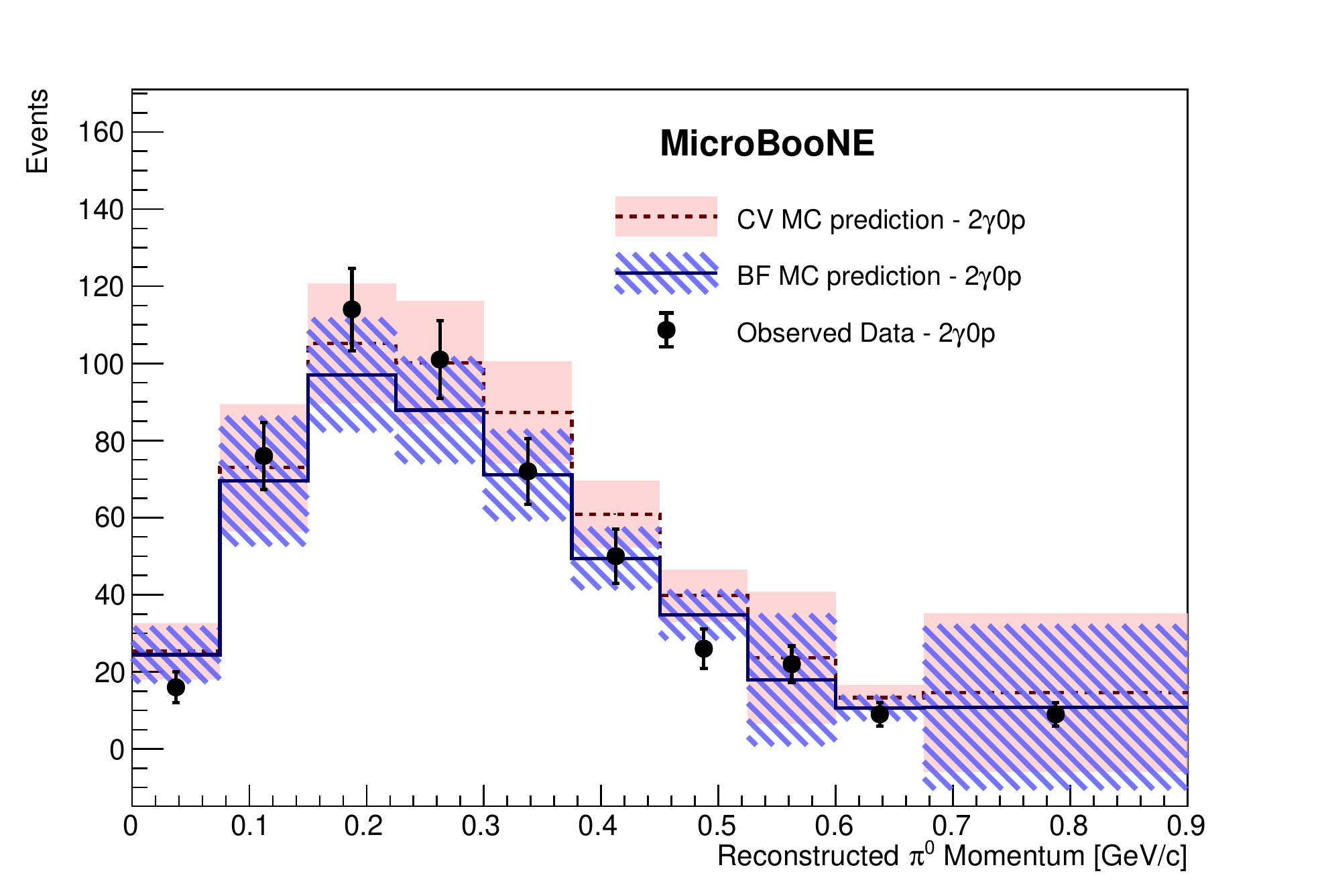}
  \caption{$2\gamma0p$}
  \end{subfigure}
  \caption{The data-MC comparison for (a) $2\gamma1p$ and  (b) $2\gamma0p$ selections, as a function of reconstructed $\pi^0$ momentum. Monte Carlo predictions at the central value and at the best-fit point ($\text{N}_{coh} = 2.6$, $a=0.98$, $b=-1.0$~[c/GeV]) are both shown, with prediction and corresponding systematic error evaluated at the \textsc{genie} central value in salmon, and at the best-fit in blue. Note that the systematic uncertainties on the plot include MC intrinsic statistical error and all the systematic errors (flux, cross-section and detector), with the exception of cross section normalization uncertainties on coherent and non-coherent NC 1$\pi^0$.}
  \label{fig:momentum_BF}
\end{figure}

The data-extracted best-fit parameters correspond to $a=0.98$ and $b=-1.0$~[c/GeV] for the scaling parameters of the non-coherent NC $1\pi^0$ events, and $N_{coh}=2.6$ for the NC coherent $\pi^0$ normalization factor with no enhancement, $N_{coh}=1$, being allowed within the $1\sigma$ error bands. This best-fit gives a $\chi^2$ per degree of freedom ($dof$) of 8.46/17. The $\chi^2/dof$ at the \textsc{genie} central value (CV) prediction is 13.74/20 yielding a $\Delta\chi^2$ between the \textsc{genie} CV and the best-fit point of 5.28 for 3 $dof$. Although the goodness-of-fit $\chi^2/dof$ values for both scenarios are acceptable due to the generally large uncertainties, the momentum-dependent shift is preferred over the \textsc{genie} CV at the 1.43$\sigma$ level. The 1D marginalized $\Delta\chi^2$ distributions in Fig.~\ref{fig:marginalized_chi} also confirm that the \textsc{genie} CV prediction agrees with data within uncertainty. The data and MC comparisons of the reconstructed $\pi^0$ momentum distributions scaled to the best-fit parameters are provided in Fig.~\ref{fig:momentum_BF} and, compared to those corresponding to the \textsc{genie} CV, show better agreement with data after the fit.

While the data suggest that \textsc{genie} may over-estimate NC $1\pi^0$ production, the results demonstrate that the \textsc{genie} prediction of NC $1\pi^{0}$s is accurate within uncertainty. This validates the approach of using the measured NC $1\pi^{0}$ event rate as a powerful \textit{in situ} constraint of \textsc{genie}-predicted NC $1\pi^0$ backgrounds as in \cite{MicroBooNE:2021zai}.  We stress that while this result motivates a momentum dependent shift in how we model our NC $\pi^0$ events, we do not apply this change to our modeling, relying instead on the fact that the assigned uncertainty covers the observed discrepancy. Future work will investigate further the possibility of tuning \textsc{genie} with results such as these to obtain better model predictions. On the other hand, it is natural to extract a data-driven NC 1$\pi^{0}$ cross-section on argon using these selections, and compare to a number of neutrino event generators, including \textsc{genie}. This is described below.

\section{Inclusive and Exclusive NC 1$\pi^0$ Cross-Sections on Argon}

\subsection{Methodology} 

The prescription for calculating the cross section is provided in Eq.~(\ref{eqn:xsec}) where the components are defined as follows: $N^\text{obs}_{NC1\pi^0}$, $N_\text{cosmic}$, and $N_{\text{bkg}}$ denote the number of selected data events, the number of background events arising from cosmic rays traversing the detector, and the number of expected beam-correlated background events, respectively; $\epsilon_{NC1\pi^0}$ denotes the efficiency of selecting NC1$\pi^0$ events; $\Phi$ denotes the integrated flux; and $N_{\text{targets}}$ denotes the number of argon atoms in the fiducial volume of the analysis.

\begin{equation}
\label{eqn:xsec}
    \sigma_{NC1\pi^{0}} = \frac{N^\text{obs}_{NC1\pi^{0}}-N_{\text{cosmic}}-N_{\text{bkg}}}{\epsilon_{NC1\pi^{0}}\Phi N_{\text{targets}}}.
\end{equation} 

This calculation is performed independently using each of the 2$\gamma1p$ and 2$\gamma0p$ selections to measure an exclusive cross section. These measurements are denoted as the NC$\pi^{0}$+1p and NC$\pi^{0}$+0p cross sections, respectively; in each case one or zero protons is explicitly required in the signal definition (described in detail below). Additionally, the calculation is performed using the combined 2$\gamma(0+1)p$ selection to measure a semi-inclusive cross section, NC$\pi^{0}$, with no requirement on the number of protons in the signal definition. Note that this semi-inclusive measurement is efficiency-corrected to include 2+ proton final states that are not included in the final selected events (11\% of the total number of NC $\pi^0$ interactions, per \textsc{GENIE}). As noted in Section \ref{sec:systematics}, the simulation is run multiple times to encompass the effect of varying underlying sources of systematic uncertainty. The calculation of each cross section is performed separately in each of these systematic ``universes'' to guarantee that all correlations between components of the cross section are handled correctly. This is done using tools from the MINERvA Analysis Toolkit \cite{MINERvA:2021ddh}.

Both selections, as well as their combination, correspond to approximately, but not identically, the same POT, provided in Table~\ref{tab:summary_tab} (due to differences in the computational processing of the two samples). To extract the semi-inclusive cross section from the combined 2$\gamma(0+1)p$ selection, the relevant 2$\gamma0p$ distributions are scaled down by the ratio between the POT of the 2$\gamma1p$ data sample (smaller POT) and the POT of the 2$\gamma0p$ data sample and then are added to the 2$\gamma1p$ distributions. This operation is performed for $N^\text{obs}_{NC1\pi^0}$, $N_\text{cosmic}$, $N_{bkg}$, and the numerator of the efficiency. 

$N^\text{obs}_{NC1\pi^0}$ and $N_\text{cosmic}$ are measured in data and therefore there is no systematic uncertainty attributed to them. These values are reported in Table~\ref{tab:summary_tab}. $N_{bkg}$ is extracted from the simulation, and we note that many of the key backgrounds in this analysis are shared with MicroBooNE's search for NC $\Delta$ radiative decay \cite{MicroBooNE:2021zai}. The dominant contributions to the uncertainty on the background event rate for each analysis are from FSI related to inelastic nucleon scattering, pion and nucleon absorption, and pion charge-exchange. The axial and vector mass parameters, $m_A$ and $m_V$, respectively, in the charged current resonant form factors are also sources of significant uncertainties; this is consistent with expectation because of the large background due to charged-current interactions in which a $\pi^0$ is produced. 

The efficiency of the selection is constructed using as the numerator the number of signal events passing all reconstruction cuts and analysis BDTs in simulation and as the denominator the total number of signal events preceding the application of any cuts or analysis BDTs. The difference in signal definition between the semi-inclusive measurement and each of the two exclusive measurements is contained in the efficiency denominator. The exclusive measurements and the semi-inclusive measurement each use a distinct efficiency denominator, reflecting the total number of simulated events truly satisfying the corresponding signal definition. In each of the exclusive measurements, the signal definition is taken to be NC1$\pi^{0}$ with exactly zero or one final-state proton with a kinetic energy above 50 MeV. In the semi-inclusive measurement, the signal definition is taken to be NC1$\pi^{0}$, notably allowing for any number of protons in the final state. The efficiency for each analysis is reported in Table~\ref{tab:summary_tab}.

The integrated flux is calculated separately for all four neutrino species ($\nu_{\mu}$, $\bar{\nu}_{\mu}$, $\nu_{e}$, $\bar{\nu}_{e}$), and the sum of these integrated fluxes is used to normalize each cross section measurement. This choice was made because of the inability to identify the species of the incident neutrino based on the neutral current final state. The integrated flux is varied within each flux systematic ``universe'', and the correlations between each varied flux and the corresponding variations in the predicted background and efficiency are taken into account when extracting the cross sections.

The number of argon atoms used is calculated as $N_{\text{targets}} = \rho V N_A/M_{Ar}$, where $V = 5.64\times10^{7} \text{cm}^3$ is the fiducial volume of the analysis, $\rho = 1.3954$~g/cm$^3$ is the density of argon at the temperature in the cryostat, and $M_{Ar} = 39.948$~g/mol is the molar mass of argon. A 1\% uncertainty is assigned to the number of targets to reflect variation in the argon density through temperature and pressure fluctuations.

\begin{table*}[ht!]
    \caption{Summary table of all inputs to the cross section calculation, reported as  $\sigma \pm\text{sys}\pm \text{stat}$ uncertainty. Note that while the individual errors on the components are given here, the full uncertainty on the cross section is calculated properly assuming full correlations.}
    \begin{adjustwidth}{-1.25cm}{-1cm}
    \centering
    \setlength{\tabcolsep}{2pt}
    \begin{tabular}{|l|c|c|c|c|}
    \hline\hline
     \textbf{} & \textbf{NC$\pi^0$ (semi-inclusive)} & \textbf{NC$\pi^0+1 p$ (exclusive)} & \textbf{NC$\pi^0+0 p$ (exclusive)}  \\ \hline\hline
    \textbf{Samples Used} & {$2\gamma(0+1)p$ Selection } & $2\gamma1p$ Selection & $2\gamma0p$ Selection \\ \hline\hline

     N$_\text{targets}$ [$10^{30}$ Ar atoms]                    &
     \multicolumn{3}{c|}{1.187 $\pm$ 0.119 $\pm$ 0.00}                                                               \\
     Flux [$10^{-10}$ $\nu$/POT/cm$^{2}$]               & \multicolumn{3}{c|}{7.876 $\pm$ 0.902 $\pm$ 0.00}                                                               \\ \cline{2-4}
     POT of sample [$10^{20}$ POT]                      & 5.84    $\pm$ 0.12    $\pm$ 0.00   & 5.84     $\pm$ 0.12    $\pm$ 0.0      & 5.89    $\pm$ 0.12   $\pm$ 0.00    \\
	 Efficiency                                         & 0.089   $\pm$ 0.003   $\pm$ 0.001  & 0.107    $\pm$ 0.006   $\pm$ 0.002    & 0.060   $\pm$ 0.003  $\pm$ 0.001   \\
	 Selected data [evts]                              	& 1125.9  $\pm$ 0.0     $\pm$ 33.5   & 634.0    $\pm$ 0.0     $\pm$ 25.2     & 496.0   $\pm$ 0.0    $\pm$ 22.3    \\
	 Cosmic data [evts]                                      & 177.0   $\pm$ 0.0     $\pm$ 8.9    & 96.1     $\pm$ 0.0     $\pm$ 6.5      & 81.5    $\pm$ 0.0    $\pm$ 6.1     \\
	 Background [evts]                                         & 345.8   $\pm$ 51.1    $\pm$ 9.0    & 279.6    $\pm$ 43.5    $\pm$ 7.2      & 208.3   $\pm$ 33.5   $\pm$ 7.0     \\
	 Background-subtracted rate [evts]                        	& 603.2   $\pm$ 51.1    $\pm$ 35.8   & 258.3    $\pm$ 43.5    $\pm$ 27.0     & 206.1   $\pm$ 33.5   $\pm$ 24.1    \\ \hline
	 $\sigma_{NC1\pi^0}$ [$10^{-38}$cm$^2$/Ar]          & 1.243   $\pm$ 0.185   $\pm$ 0.076  & 0.444    $\pm$ 0.098   $\pm$ 0.047    & 0.624   $\pm$ 0.131  $\pm$ 0.075   \\
    \hline\hline
    \end{tabular}
    \end{adjustwidth}
    \label{tab:summary_tab}
\end{table*}

\subsection{Results and Interpretation}

\begin{figure}[h]
	\centering
	\includegraphics[width=0.49\textwidth]{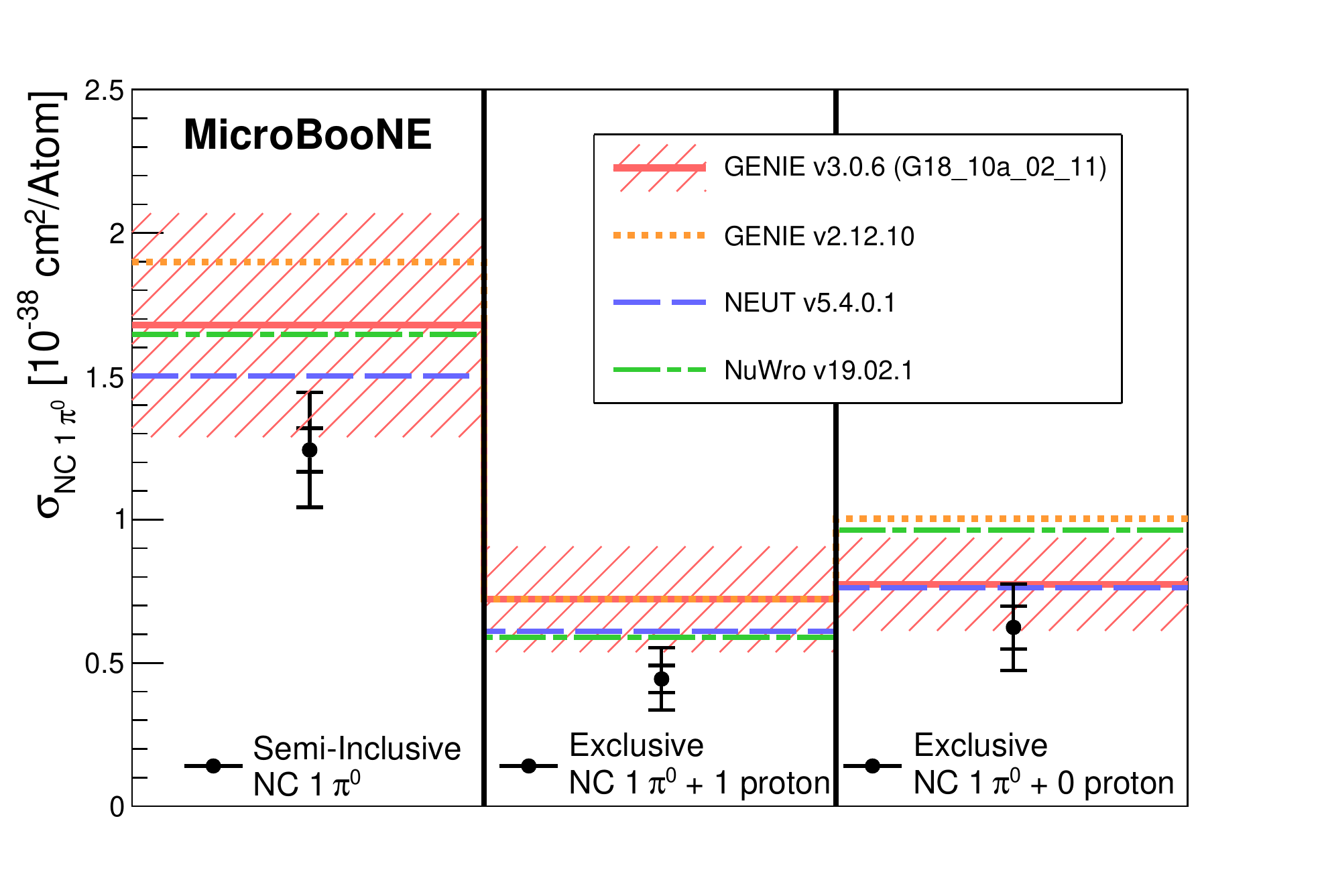}
    \caption{Measured semi-inclusive NC$\pi^0$, exclusive NC$\pi^0$+1p, and exclusive NC$\pi^0$+0p cross sections, each compared to the corresponding \textsc{genie} v3 (G18\_10a\_02\_11) cross section and its uncertainty (shaded red bands) as well as other contemporary neutrino generators. Inner error bars on data points are statistical only; outer are statistical and systematic, summed in quadrature.}
	\label{fig:xsec_comparison}
\end{figure}

The calculation of each cross section from its components follows from Eq.~(\ref{eqn:xsec}) and is summarized in Table~\ref{tab:summary_tab}. The resulting cross sections are shown in Fig.~\ref{fig:xsec_comparison}, compared to the simulated cross sections from several neutrino event generators including \textsc{genie}, \textsc{NuWro} \cite{Golan:2012rfa}, and \textsc{neut} \cite{Hayato:2002sd}. \textsc{NEUT} and \textsc{GENIE} both use the Berger-Sehgal model as their foundation for modeling pion production in the $\Delta(1232)$ resonance region, while \textsc{NuWro} implements a custom model optimized for the $\Delta$ resonance peak region. The complete details of the models used in each generator are discussed at length in Ref.~\cite{Avanzini:2021qlx}. The \textsc{genie} curve shown is generated using the MicroBooNE cross-section ``tune'' \cite{MicroBooNE:2021ccs}, which does not modify the \textsc{genie} v3.0.6 central value prediction (because the tune did not adjust the NC interaction model), but does define the uncertainty on the prediction. The error bars on the data points include systematic error associated with the modeling of background events that enters into the cross-sections via background subtraction as well as error associated with the modeling of the signal events that enters into the cross-sections via efficiency correction. The shaded band around the \textsc{GENIE} central value prediction shows the error associated with the prediction of the signal cross section. 

We observe a consistent deficit in data compared to \textsc{genie} for the combined semi-inclusive measurement and for each of the individual NC$\pi^{0}$+1p and NC$\pi^{0}$+0p exclusive measurements. Overall, the \textsc{neut} predictions most closely match the reported measurements across semi-inclusive and exclusive final states. Additionally we note that while \textsc{NuWro} is generally consistent with the other generators in its semi-inclusive and exclusive 1p predictions, its exclusive 0p prediction is higher compared to \textsc{neut} and \textsc{genie} predictions. The extracted semi-inclusive NC$\pi^{0}$ cross section is 1.24 $\pm$ 0.19 (syst) $\pm$ 0.08 (stat) [$10^{-38}$cm$^2$/Ar] which is 26\% lower than the \textsc{genie} prediction of 1.68 [$10^{-38}$cm$^2$/Ar]. We calculate a $\chi^2$ test statistic comparing our data measurement with full uncertainties to the CV of each model prediction for both exclusive NC$\pi^0$+0p and NC$\pi^0$+1p cross sections simultaneously (\textit{i.e.} two degrees of freedom). The resulting values are 7.6, 7.7, 2.4, and 5.1 for comparisons against \textsc{genie} v3, \textsc{genie} v2, \textsc{neut}, and \textsc{NuWro}, respectively.

\begin{figure}[h]
	\centering
	\includegraphics[width=0.49\textwidth]{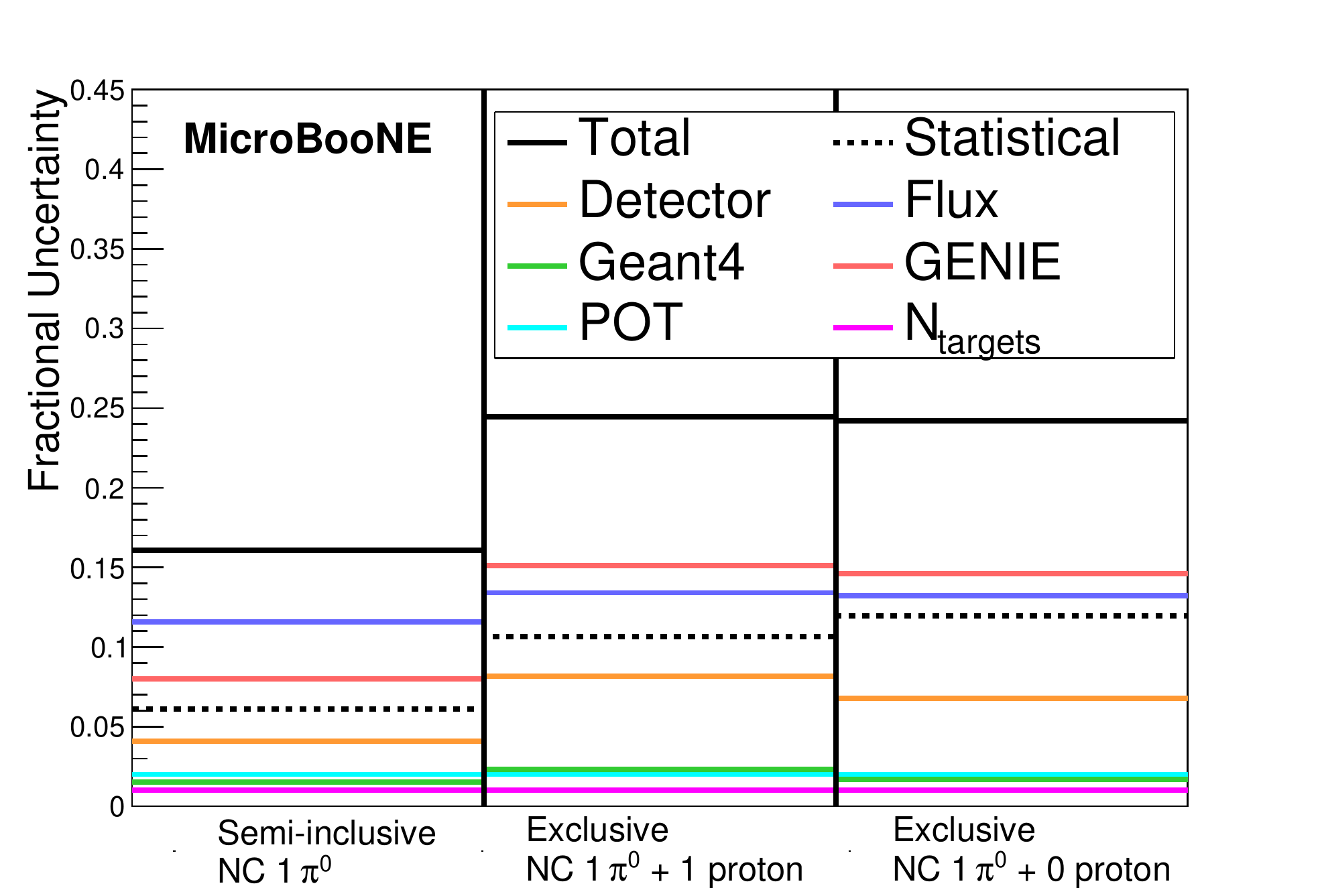}
     \caption{Error budget for the semi-inclusive NC$\pi^0$, exclusive NC$\pi^0$+1p, and exclusive NC$\pi^0$+0p cross section measurements.}
	\label{fig:xsec_uncertainty}
\end{figure}

The corresponding breakdown of uncertainty for each of the measurement channels is shown in Fig.~\ref{fig:xsec_uncertainty}. In all cases the flux, \textsc{genie}, and statistical uncertainties are dominant. The dominant contributions to the \textsc{genie} uncertainties enter into the cross section via the background subtraction and, as noted above, arise from the modeling of final-state interactions and the axial and vector mass parameters governing CC resonant pion production.
 
\begin{figure}[h]
	\centering
	\includegraphics[width=0.49\textwidth]{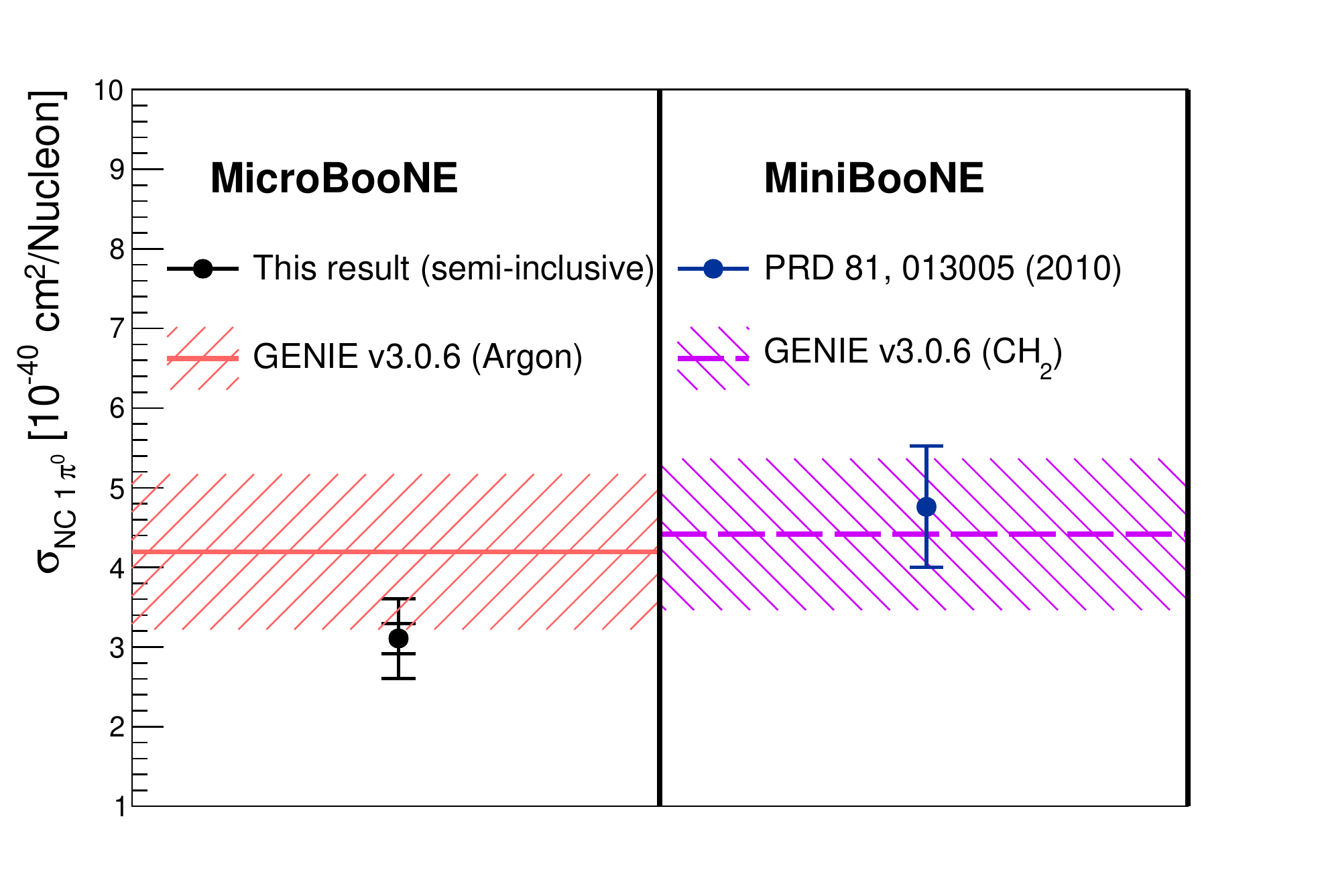}
    \caption{Comparison of this semi-inclusive result on argon (left) as well as that from MiniBooNE on mineral oil CH$_2$ (right), to the same \textsc{GENIE} v3.0.6. While the published MiniBooNE result was originally compared to a prediction made using the \textsc{NUANCE} v3 generator \cite{Casper:2002sd}, we have instead generated a prediction using GENIE to aid in a comparison between the two experimental results. MiniBooNE's statistical uncertainty is small and only the systematic error bar is visible. Shaded error bands show \textsc{GENIE} uncertainty only.}
	\label{fig:compare_uB}
\end{figure}

To further understand this measurement, it is instructive to compare it to previous experimental measurements of NC$\pi^0$ production. We compare our measurement to that performed by MiniBooNE which operated in the same beamline as MicroBooNE but which utilized a different detector material (mineral oil, CH$_2$) as the neutrino scattering target. In MiniBooNE's  NC $\pi^0$ analysis, they measured NC interactions wherein only one $\pi^0$ and no additional mesons exited the target nucleus (no requirement on the number or identity of outgoing nucleons was made). A final flux-averaged cross section of 4.76 $\pm$ 0.76 $\pm$ 0.05 [$10^{-40}$cm$^2$/nucleon] was reported \cite{MiniBooNE:2009dxl}. We can compare this result to our semi-inclusive result by comparing each to the same neutrino generator. This is shown in Fig.~\ref{fig:compare_uB} where we compare both to the default \textsc{GENIE} v3.0.6 on argon and mineral oil respectively. We observe that while this result on argon lies slightly below the expected central value, both our result and MiniBooNE's agree with \textsc{GENIE} v3.0.6 within assigned uncertainties. 

\section{Summary}

In summary, we report the highest statistics measurement to date of neutrino neutral current single pion production on argon, including the first exclusive measurements of this process ever made in argon. These cross sections are measured using the MicroBooNE detector exposed to the Fermilab Booster Neutrino Beamline, which has $\langle E_{\nu} \rangle <1$ GeV. As presented within this paper,  kinematic distributions of the $\pi^0$ momentum and angle relative to the beam direction provide some sensitivity to contributions to this process from coherent and non-coherent pion production and suggest that, given the currently analyzed MicroBooNE data statistics, the nominal \textsc{genie} neutrino event generator used for MicroBooNE Monte Carlo modeling describes the observed distributions within uncertainties. This has provided an important validation check justifying the use of this sample as a powerful constraint for backgrounds to single-photon searches in MicroBooNE, \textit{e.g.}~in \cite{MicroBooNE:2021zai}. 

Using a total of 1,130 observed NC $\pi^0$ events, a flux-averaged cross section has been extracted for neutrinos with a mean energy of 804 MeV and has been found to correspond to $1.243\pm0.185$ (syst) $\pm0.076$ (stat), $0.444\pm0.098\pm0.047$, and $0.624\pm0.131\pm0.075$ $[10^{-38}\textrm{cm}^2/\textrm{Ar}]$ for the semi-inclusive NC$\pi^0$, exclusive NC$\pi^0$+1p, and exclusive NC$\pi^0$+0p processes compared to $1.678$, $0.722$, and $0.774$ $[10^{-38}\textrm{cm}^2/\textrm{Ar}]$ in the default \textsc{genie} prediction used by MicroBooNE. Comparison to other generators including \textsc{neut} and \textsc{NuWro} show reasonable agreement with the \textsc{neut} predictions found to be slightly more consistent with the MicroBooNE data-extracted cross-section for all three exclusive and semi-inclusive processes.

\vskip 0.2in

\begin{acknowledgments}
This document was prepared by the MicroBooNE collaboration using the resources of the Fermi National Accelerator Laboratory (Fermilab), a U.S. Department of Energy, Office of Science, HEP User Facility. Fermilab is managed by Fermi Research Alliance, LLC (FRA), acting under Contract No. DE-AC02-07CH11359.  MicroBooNE is supported by the following: the U.S. Department of Energy, Office of Science, Offices of High Energy Physics and Nuclear Physics; the U.S. National Science Foundation; the Swiss National Science Foundation; the Science and Technology Facilities Council (STFC), part of the United Kingdom Research and Innovation; the Royal Society (United Kingdom); and The European Union’s Horizon 2020 Marie Sklodowska-Curie Actions. Additional support for the laser calibration system and cosmic ray tagger was provided by the Albert Einstein Center for Fundamental Physics, Bern, Switzerland. We also acknowledge the contributions of technical and scientific staff to the design, construction, and operation of the MicroBooNE detector as well as the contributions of past collaborators to the development of MicroBooNE analyses, without whom this work would not have been possible.
\end{acknowledgments}

\bibliographystyle{apsrev4-1} 
\bibliography{main}

\begin{thebibliography}{56}%
\makeatletter
\providecommand \@ifxundefined [1]{%
 \@ifx{#1\undefined}
}%
\providecommand \@ifnum [1]{%
 \ifnum #1\expandafter \@firstoftwo
 \else \expandafter \@secondoftwo
 \fi
}%
\providecommand \@ifx [1]{%
 \ifx #1\expandafter \@firstoftwo
 \else \expandafter \@secondoftwo
 \fi
}%
\providecommand \natexlab [1]{#1}%
\providecommand \enquote  [1]{``#1''}%
\providecommand \bibnamefont  [1]{#1}%
\providecommand \bibfnamefont [1]{#1}%
\providecommand \citenamefont [1]{#1}%
\providecommand \href@noop [0]{\@secondoftwo}%
\providecommand \href [0]{\begingroup \@sanitize@url \@href}%
\providecommand \@href[1]{\@@startlink{#1}\@@href}%
\providecommand \@@href[1]{\endgroup#1\@@endlink}%
\providecommand \@sanitize@url [0]{\catcode `\\12\catcode `\$12\catcode
  `\&12\catcode `\#12\catcode `\^12\catcode `\_12\catcode `\%12\relax}%
\providecommand \@@startlink[1]{}%
\providecommand \@@endlink[0]{}%
\providecommand \url  [0]{\begingroup\@sanitize@url \@url }%
\providecommand \@url [1]{\endgroup\@href {#1}{\urlprefix }}%
\providecommand \urlprefix  [0]{URL }%
\providecommand \Eprint [0]{\href }%
\providecommand \doibase [0]{http://dx.doi.org/}%
\providecommand \selectlanguage [0]{\@gobble}%
\providecommand \bibinfo  [0]{\@secondoftwo}%
\providecommand \bibfield  [0]{\@secondoftwo}%
\providecommand \translation [1]{[#1]}%
\providecommand \BibitemOpen [0]{}%
\providecommand \bibitemStop [0]{}%
\providecommand \bibitemNoStop [0]{.\EOS\space}%
\providecommand \EOS [0]{\spacefactor3000\relax}%
\providecommand \BibitemShut  [1]{\csname bibitem#1\endcsname}%
\let\auto@bib@innerbib\@empty
\bibitem [{\citenamefont {Benhar}\ \emph {et~al.}(2017)\citenamefont {Benhar},
  \citenamefont {Huber}, \citenamefont {Mariani},\ and\ \citenamefont
  {Meloni}}]{Benhar:2015wva}%
  \BibitemOpen
  \bibfield  {author} {\bibinfo {author} {\bibfnamefont {O.}~\bibnamefont
  {Benhar}}, \bibinfo {author} {\bibfnamefont {P.}~\bibnamefont {Huber}},
  \bibinfo {author} {\bibfnamefont {C.}~\bibnamefont {Mariani}}, \ and\
  \bibinfo {author} {\bibfnamefont {D.}~\bibnamefont {Meloni}},\ }\href
  {\doibase 10.1016/j.physrep.2017.07.004} {\bibfield  {journal} {\bibinfo
  {journal} {Phys. Rept.}\ }\textbf {\bibinfo {volume} {700}},\ \bibinfo
  {pages} {1} (\bibinfo {year} {2017})},\ \Eprint
  {http://arxiv.org/abs/1501.06448} {arXiv:1501.06448 [nucl-th]} \BibitemShut
  {NoStop}%
\bibitem [{\citenamefont {Antonello}\ \emph {et~al.}(2015)\citenamefont
  {Antonello} \emph {et~al.}}]{SBNproposal}%
  \BibitemOpen
  \bibfield  {author} {\bibinfo {author} {\bibfnamefont {M.}~\bibnamefont
  {Antonello}} \emph {et~al.} (\bibinfo {collaboration} {MicroBooNE, LAr1-ND,
  ICARUS-WA104}),\ }\href@noop {} {\  (\bibinfo {year} {2015})},\ \Eprint
  {http://arxiv.org/abs/1503.01520} {arXiv:1503.01520 [physics.ins-det]}
  \BibitemShut {NoStop}%
\bibitem [{\citenamefont {Abi}\ \emph {et~al.}(2020)\citenamefont {Abi} \emph
  {et~al.}}]{DUNE:2020ypp}%
  \BibitemOpen
  \bibfield  {author} {\bibinfo {author} {\bibfnamefont {B.}~\bibnamefont
  {Abi}} \emph {et~al.} (\bibinfo {collaboration} {DUNE}),\ }\href@noop {} {\
  (\bibinfo {year} {2020})},\ \Eprint {http://arxiv.org/abs/2002.03005}
  {arXiv:2002.03005 [hep-ex]} \BibitemShut {NoStop}%
\bibitem [{\citenamefont {Abratenko}\ \emph {et~al.}(2022)\citenamefont
  {Abratenko} \emph {et~al.}}]{MicroBooNE:2021zai}%
  \BibitemOpen
  \bibfield  {author} {\bibinfo {author} {\bibfnamefont {P.}~\bibnamefont
  {Abratenko}} \emph {et~al.} (\bibinfo {collaboration} {MicroBooNE}),\ }\href
  {\doibase 10.1103/PhysRevLett.128.111801} {\bibfield  {journal} {\bibinfo
  {journal} {Phys. Rev. Lett.}\ }\textbf {\bibinfo {volume} {128}},\ \bibinfo
  {pages} {111801} (\bibinfo {year} {2022})},\ \Eprint
  {http://arxiv.org/abs/2110.00409} {arXiv:2110.00409 [hep-ex]} \BibitemShut
  {NoStop}%
\bibitem [{\citenamefont {Bertuzzo}\ \emph {et~al.}(2018)\citenamefont
  {Bertuzzo}, \citenamefont {Jana}, \citenamefont {Machado},\ and\
  \citenamefont {Zukanovich~Funchal}}]{Bertuzzo:2018itn}%
  \BibitemOpen
  \bibfield  {author} {\bibinfo {author} {\bibfnamefont {E.}~\bibnamefont
  {Bertuzzo}}, \bibinfo {author} {\bibfnamefont {S.}~\bibnamefont {Jana}},
  \bibinfo {author} {\bibfnamefont {P.~A.~N.}\ \bibnamefont {Machado}}, \ and\
  \bibinfo {author} {\bibfnamefont {R.}~\bibnamefont {Zukanovich~Funchal}},\
  }\href {\doibase 10.1103/PhysRevLett.121.241801} {\bibfield  {journal}
  {\bibinfo  {journal} {Phys. Rev. Lett.}\ }\textbf {\bibinfo {volume} {121}},\
  \bibinfo {pages} {241801} (\bibinfo {year} {2018})},\ \Eprint
  {http://arxiv.org/abs/1807.09877} {arXiv:1807.09877 [hep-ph]} \BibitemShut
  {NoStop}%
\bibitem [{\citenamefont {Ballett}\ \emph {et~al.}(2019)\citenamefont
  {Ballett}, \citenamefont {Pascoli},\ and\ \citenamefont
  {Ross-Lonergan}}]{Ballett:2018ynz}%
  \BibitemOpen
  \bibfield  {author} {\bibinfo {author} {\bibfnamefont {P.}~\bibnamefont
  {Ballett}}, \bibinfo {author} {\bibfnamefont {S.}~\bibnamefont {Pascoli}}, \
  and\ \bibinfo {author} {\bibfnamefont {M.}~\bibnamefont {Ross-Lonergan}},\
  }\href {\doibase 10.1103/PhysRevD.99.071701} {\bibfield  {journal} {\bibinfo
  {journal} {Phys. Rev. D}\ }\textbf {\bibinfo {volume} {99}},\ \bibinfo
  {pages} {071701} (\bibinfo {year} {2019})},\ \Eprint
  {http://arxiv.org/abs/1808.02915} {arXiv:1808.02915 [hep-ph]} \BibitemShut
  {NoStop}%
\bibitem [{\citenamefont {Abdullahi}\ \emph {et~al.}(2021)\citenamefont
  {Abdullahi}, \citenamefont {Hostert},\ and\ \citenamefont
  {Pascoli}}]{Abdullahi:2020nyr}%
  \BibitemOpen
  \bibfield  {author} {\bibinfo {author} {\bibfnamefont {A.}~\bibnamefont
  {Abdullahi}}, \bibinfo {author} {\bibfnamefont {M.}~\bibnamefont {Hostert}},
  \ and\ \bibinfo {author} {\bibfnamefont {S.}~\bibnamefont {Pascoli}},\ }\href
  {\doibase 10.1016/j.physletb.2021.136531} {\bibfield  {journal} {\bibinfo
  {journal} {Phys. Lett. B}\ }\textbf {\bibinfo {volume} {820}},\ \bibinfo
  {pages} {136531} (\bibinfo {year} {2021})},\ \Eprint
  {http://arxiv.org/abs/2007.11813} {arXiv:2007.11813 [hep-ph]} \BibitemShut
  {NoStop}%
\bibitem [{\citenamefont {Dutta}\ \emph {et~al.}(2020)\citenamefont {Dutta},
  \citenamefont {Ghosh},\ and\ \citenamefont {Li}}]{Dutta:2020scq}%
  \BibitemOpen
  \bibfield  {author} {\bibinfo {author} {\bibfnamefont {B.}~\bibnamefont
  {Dutta}}, \bibinfo {author} {\bibfnamefont {S.}~\bibnamefont {Ghosh}}, \ and\
  \bibinfo {author} {\bibfnamefont {T.}~\bibnamefont {Li}},\ }\href {\doibase
  10.1103/PhysRevD.102.055017} {\bibfield  {journal} {\bibinfo  {journal}
  {Phys. Rev. D}\ }\textbf {\bibinfo {volume} {102}},\ \bibinfo {pages}
  {055017} (\bibinfo {year} {2020})},\ \Eprint
  {http://arxiv.org/abs/2006.01319} {arXiv:2006.01319 [hep-ph]} \BibitemShut
  {NoStop}%
\bibitem [{\citenamefont {Abdallah}\ \emph {et~al.}(2021)\citenamefont
  {Abdallah}, \citenamefont {Gandhi},\ and\ \citenamefont
  {Roy}}]{Abdallah:2020vgg}%
  \BibitemOpen
  \bibfield  {author} {\bibinfo {author} {\bibfnamefont {W.}~\bibnamefont
  {Abdallah}}, \bibinfo {author} {\bibfnamefont {R.}~\bibnamefont {Gandhi}}, \
  and\ \bibinfo {author} {\bibfnamefont {S.}~\bibnamefont {Roy}},\ }\href
  {\doibase 10.1103/PhysRevD.104.055028} {\bibfield  {journal} {\bibinfo
  {journal} {Phys. Rev. D}\ }\textbf {\bibinfo {volume} {104}},\ \bibinfo
  {pages} {055028} (\bibinfo {year} {2021})},\ \Eprint
  {http://arxiv.org/abs/2010.06159} {arXiv:2010.06159 [hep-ph]} \BibitemShut
  {NoStop}%
\bibitem [{\citenamefont {Cianci}\ \emph {et~al.}(2017)\citenamefont {Cianci},
  \citenamefont {Furmanski}, \citenamefont {Karagiorgi},\ and\ \citenamefont
  {Ross-Lonergan}}]{Cianci:2017okw}%
  \BibitemOpen
  \bibfield  {author} {\bibinfo {author} {\bibfnamefont {D.}~\bibnamefont
  {Cianci}}, \bibinfo {author} {\bibfnamefont {A.}~\bibnamefont {Furmanski}},
  \bibinfo {author} {\bibfnamefont {G.}~\bibnamefont {Karagiorgi}}, \ and\
  \bibinfo {author} {\bibfnamefont {M.}~\bibnamefont {Ross-Lonergan}},\ }\href
  {\doibase 10.1103/PhysRevD.96.055001} {\bibfield  {journal} {\bibinfo
  {journal} {Phys. Rev. D}\ }\textbf {\bibinfo {volume} {96}},\ \bibinfo
  {pages} {055001} (\bibinfo {year} {2017})},\ \Eprint
  {http://arxiv.org/abs/1702.01758} {arXiv:1702.01758 [hep-ph]} \BibitemShut
  {NoStop}%
\bibitem [{\citenamefont {Furmanski}\ and\ \citenamefont
  {Hilgenberg}(2021)}]{Furmanski:2020smg}%
  \BibitemOpen
  \bibfield  {author} {\bibinfo {author} {\bibfnamefont {A.~P.}\ \bibnamefont
  {Furmanski}}\ and\ \bibinfo {author} {\bibfnamefont {C.}~\bibnamefont
  {Hilgenberg}},\ }\href {\doibase 10.1103/PhysRevD.103.112011} {\bibfield
  {journal} {\bibinfo  {journal} {Phys. Rev. D}\ }\textbf {\bibinfo {volume}
  {103}},\ \bibinfo {pages} {112011} (\bibinfo {year} {2021})},\ \Eprint
  {http://arxiv.org/abs/2012.09788} {arXiv:2012.09788 [hep-ex]} \BibitemShut
  {NoStop}%
\bibitem [{\citenamefont {Acciarri}\ \emph
  {et~al.}(2017{\natexlab{a}})\citenamefont {Acciarri} \emph
  {et~al.}}]{MicroBooNE:2016pwy}%
  \BibitemOpen
  \bibfield  {author} {\bibinfo {author} {\bibfnamefont {R.}~\bibnamefont
  {Acciarri}} \emph {et~al.} (\bibinfo {collaboration} {MicroBooNE}),\ }\href
  {\doibase 10.1088/1748-0221/12/02/P02017} {\bibfield  {journal} {\bibinfo
  {journal} {JINST}\ }\textbf {\bibinfo {volume} {12}},\ \bibinfo {pages}
  {P02017} (\bibinfo {year} {2017}{\natexlab{a}})},\ \Eprint
  {http://arxiv.org/abs/1612.05824} {arXiv:1612.05824 [physics.ins-det]}
  \BibitemShut {NoStop}%
\bibitem [{\citenamefont {Aguilar-Arevalo}\ \emph {et~al.}(2009)\citenamefont
  {Aguilar-Arevalo} \emph {et~al.}}]{BNBflux}%
  \BibitemOpen
  \bibfield  {author} {\bibinfo {author} {\bibfnamefont {A.~A.}\ \bibnamefont
  {Aguilar-Arevalo}} \emph {et~al.} (\bibinfo {collaboration} {MiniBooNE}),\
  }\href {\doibase 10.1103/PhysRevD.79.072002} {\bibfield  {journal} {\bibinfo
  {journal} {Phys. Rev. D}\ }\textbf {\bibinfo {volume} {79}},\ \bibinfo
  {pages} {072002} (\bibinfo {year} {2009})}\BibitemShut {NoStop}%
\bibitem [{\citenamefont {Aliaga}\ \emph {et~al.}(2016)\citenamefont {Aliaga}
  \emph {et~al.}}]{MINERvA:2016iqn}%
  \BibitemOpen
  \bibfield  {author} {\bibinfo {author} {\bibfnamefont {L.}~\bibnamefont
  {Aliaga}} \emph {et~al.} (\bibinfo {collaboration} {MINERvA}),\ }\href
  {\doibase 10.1103/PhysRevD.94.092005} {\bibfield  {journal} {\bibinfo
  {journal} {Phys. Rev. D}\ }\textbf {\bibinfo {volume} {94}},\ \bibinfo
  {pages} {092005} (\bibinfo {year} {2016})},\ \Eprint
  {http://arxiv.org/abs/1607.00704} {arXiv:1607.00704 [hep-ex]} \BibitemShut
  {NoStop}%
\bibitem [{\citenamefont {Acero}\ \emph {et~al.}(2020)\citenamefont {Acero}
  \emph {et~al.}}]{NOvA:2019bdw}%
  \BibitemOpen
  \bibfield  {author} {\bibinfo {author} {\bibfnamefont {M.~A.}\ \bibnamefont
  {Acero}} \emph {et~al.} (\bibinfo {collaboration} {NOvA}),\ }\href {\doibase
  10.1103/PhysRevD.102.012004} {\bibfield  {journal} {\bibinfo  {journal}
  {Phys. Rev. D}\ }\textbf {\bibinfo {volume} {102}},\ \bibinfo {pages}
  {012004} (\bibinfo {year} {2020})},\ \Eprint
  {http://arxiv.org/abs/1902.00558} {arXiv:1902.00558 [hep-ex]} \BibitemShut
  {NoStop}%
\bibitem [{\citenamefont {Abe}\ \emph {et~al.}(2013)\citenamefont {Abe} \emph
  {et~al.}}]{T2Kflux}%
  \BibitemOpen
  \bibfield  {author} {\bibinfo {author} {\bibfnamefont {K.}~\bibnamefont
  {Abe}} \emph {et~al.} (\bibinfo {collaboration} {T2K}),\ }\href {\doibase
  10.1103/PhysRevD.87.012001} {\bibfield  {journal} {\bibinfo  {journal} {Phys.
  Rev. D}\ }\textbf {\bibinfo {volume} {87}},\ \bibinfo {pages} {012001}
  (\bibinfo {year} {2013})},\ \Eprint {http://arxiv.org/abs/1211.0469}
  {arXiv:1211.0469 [hep-ex]} \BibitemShut {NoStop}%
\bibitem [{\citenamefont {Abe}\ \emph {et~al.}(2015)\citenamefont {Abe} \emph
  {et~al.}}]{Hyper-KamiokandeProto-:2015xww}%
  \BibitemOpen
  \bibfield  {author} {\bibinfo {author} {\bibfnamefont {K.}~\bibnamefont
  {Abe}} \emph {et~al.} (\bibinfo {collaboration} {Hyper-Kamiokande
  Proto-Collaboration}),\ }\href {\doibase 10.1093/ptep/ptv061} {\bibfield
  {journal} {\bibinfo  {journal} {PTEP}\ }\textbf {\bibinfo {volume} {2015}},\
  \bibinfo {pages} {053C02} (\bibinfo {year} {2015})},\ \Eprint
  {http://arxiv.org/abs/1502.05199} {arXiv:1502.05199 [hep-ex]} \BibitemShut
  {NoStop}%
\bibitem [{\citenamefont {Acciarri}\ \emph
  {et~al.}(2017{\natexlab{b}})\citenamefont {Acciarri} \emph
  {et~al.}}]{ArgoNeuT:2015ldo}%
  \BibitemOpen
  \bibfield  {author} {\bibinfo {author} {\bibfnamefont {R.}~\bibnamefont
  {Acciarri}} \emph {et~al.} (\bibinfo {collaboration} {ArgoNeuT}),\ }\href
  {\doibase 10.1103/PhysRevD.96.012006} {\bibfield  {journal} {\bibinfo
  {journal} {Phys. Rev. D}\ }\textbf {\bibinfo {volume} {96}},\ \bibinfo
  {pages} {012006} (\bibinfo {year} {2017}{\natexlab{b}})},\ \Eprint
  {http://arxiv.org/abs/1511.00941} {arXiv:1511.00941 [hep-ex]} \BibitemShut
  {NoStop}%
\bibitem [{\citenamefont {Barish}\ \emph {et~al.}(1974)\citenamefont {Barish}
  \emph {et~al.}}]{Barish:1974fe}%
  \BibitemOpen
  \bibfield  {author} {\bibinfo {author} {\bibfnamefont {S.~J.}\ \bibnamefont
  {Barish}} \emph {et~al.},\ }\href {\doibase 10.1103/PhysRevLett.33.448}
  {\bibfield  {journal} {\bibinfo  {journal} {Phys. Rev. Lett.}\ }\textbf
  {\bibinfo {volume} {33}},\ \bibinfo {pages} {448} (\bibinfo {year}
  {1974})}\BibitemShut {NoStop}%
\bibitem [{\citenamefont {Derrick}\ \emph {et~al.}(1981)\citenamefont {Derrick}
  \emph {et~al.}}]{Derrick:1980xw}%
  \BibitemOpen
  \bibfield  {author} {\bibinfo {author} {\bibfnamefont {M.}~\bibnamefont
  {Derrick}} \emph {et~al.},\ }\href {\doibase 10.1103/PhysRevD.23.569}
  {\bibfield  {journal} {\bibinfo  {journal} {Phys. Rev. D}\ }\textbf {\bibinfo
  {volume} {23}},\ \bibinfo {pages} {569} (\bibinfo {year} {1981})}\BibitemShut
  {NoStop}%
\bibitem [{\citenamefont {Lee}\ \emph {et~al.}(1977)\citenamefont {Lee} \emph
  {et~al.}}]{Lee:1976wr}%
  \BibitemOpen
  \bibfield  {author} {\bibinfo {author} {\bibfnamefont {W.-Y.}\ \bibnamefont
  {Lee}} \emph {et~al.},\ }\href {\doibase 10.1103/PhysRevLett.38.202}
  {\bibfield  {journal} {\bibinfo  {journal} {Phys. Rev. Lett.}\ }\textbf
  {\bibinfo {volume} {38}},\ \bibinfo {pages} {202} (\bibinfo {year}
  {1977})}\BibitemShut {NoStop}%
\bibitem [{\citenamefont {Krenz}\ \emph {et~al.}(1978)\citenamefont {Krenz}
  \emph {et~al.}}]{GargamelleNeutrinoPropane:1977hya}%
  \BibitemOpen
  \bibfield  {author} {\bibinfo {author} {\bibfnamefont {W.}~\bibnamefont
  {Krenz}} \emph {et~al.} (\bibinfo {collaboration} {Gargamelle Neutrino
  Propane, Aachen-Brussels-CERN-Ecole Poly-Orsay-Padua}),\ }\href {\doibase
  10.1016/0550-3213(78)90213-4} {\bibfield  {journal} {\bibinfo  {journal}
  {Nucl. Phys. B}\ }\textbf {\bibinfo {volume} {135}},\ \bibinfo {pages} {45}
  (\bibinfo {year} {1978})}\BibitemShut {NoStop}%
\bibitem [{\citenamefont {Aguilar-Arevalo}\ \emph {et~al.}(2008)\citenamefont
  {Aguilar-Arevalo} \emph {et~al.}}]{MiniBooNE:2008mmr}%
  \BibitemOpen
  \bibfield  {author} {\bibinfo {author} {\bibfnamefont {A.~A.}\ \bibnamefont
  {Aguilar-Arevalo}} \emph {et~al.} (\bibinfo {collaboration} {MiniBooNE}),\
  }\href {\doibase 10.1016/j.physletb.2008.05.006} {\bibfield  {journal}
  {\bibinfo  {journal} {Phys. Lett. B}\ }\textbf {\bibinfo {volume} {664}},\
  \bibinfo {pages} {41} (\bibinfo {year} {2008})},\ \Eprint
  {http://arxiv.org/abs/0803.3423} {arXiv:0803.3423 [hep-ex]} \BibitemShut
  {NoStop}%
\bibitem [{\citenamefont {Aguilar-Arevalo}\ \emph {et~al.}(2011)\citenamefont
  {Aguilar-Arevalo} \emph {et~al.}}]{MiniBooNE2011}%
  \BibitemOpen
  \bibfield  {author} {\bibinfo {author} {\bibfnamefont {A.~A.}\ \bibnamefont
  {Aguilar-Arevalo}} \emph {et~al.} (\bibinfo {collaboration} {MiniBooNE}),\
  }\href {\doibase 10.1103/PhysRevD.83.052009} {\bibfield  {journal} {\bibinfo
  {journal} {Phys. Rev. D}\ }\textbf {\bibinfo {volume} {83}},\ \bibinfo
  {pages} {052009} (\bibinfo {year} {2011})}\BibitemShut {NoStop}%
\bibitem [{\citenamefont {Aguilar-Arevalo}\ \emph {et~al.}(2010)\citenamefont
  {Aguilar-Arevalo} \emph {et~al.}}]{MiniBooNE:2009dxl}%
  \BibitemOpen
  \bibfield  {author} {\bibinfo {author} {\bibfnamefont {A.~A.}\ \bibnamefont
  {Aguilar-Arevalo}} \emph {et~al.} (\bibinfo {collaboration} {MiniBooNE}),\
  }\href {\doibase 10.1103/PhysRevD.81.013005} {\bibfield  {journal} {\bibinfo
  {journal} {Phys. Rev. D}\ }\textbf {\bibinfo {volume} {81}},\ \bibinfo
  {pages} {013005} (\bibinfo {year} {2010})},\ \Eprint
  {http://arxiv.org/abs/0911.2063} {arXiv:0911.2063 [hep-ex]} \BibitemShut
  {NoStop}%
\bibitem [{\citenamefont {Kurimoto}\ \emph {et~al.}(2010)\citenamefont
  {Kurimoto} \emph {et~al.}}]{SciBooNE:2010NCpi0}%
  \BibitemOpen
  \bibfield  {author} {\bibinfo {author} {\bibfnamefont {Y.}~\bibnamefont
  {Kurimoto}} \emph {et~al.} (\bibinfo {collaboration} {SciBooNE}),\ }\href
  {\doibase 10.1103/PhysRevD.81.033004} {\bibfield  {journal} {\bibinfo
  {journal} {Phys. Rev. D}\ }\textbf {\bibinfo {volume} {81}},\ \bibinfo
  {pages} {033004} (\bibinfo {year} {2010})}\BibitemShut {NoStop}%
\bibitem [{\citenamefont {Nakayama}\ \emph {et~al.}(2005)\citenamefont
  {Nakayama} \emph {et~al.}}]{K2K:2004qpv}%
  \BibitemOpen
  \bibfield  {author} {\bibinfo {author} {\bibfnamefont {S.}~\bibnamefont
  {Nakayama}} \emph {et~al.} (\bibinfo {collaboration} {K2K}),\ }\href
  {\doibase 10.1016/j.physletb.2005.05.044} {\bibfield  {journal} {\bibinfo
  {journal} {Phys. Lett. B}\ }\textbf {\bibinfo {volume} {619}},\ \bibinfo
  {pages} {255} (\bibinfo {year} {2005})},\ \Eprint
  {http://arxiv.org/abs/hep-ex/0408134} {arXiv:hep-ex/0408134} \BibitemShut
  {NoStop}%
\bibitem [{\citenamefont {Abe}\ \emph {et~al.}(2018)\citenamefont {Abe} \emph
  {et~al.}}]{T2K:2017epu}%
  \BibitemOpen
  \bibfield  {author} {\bibinfo {author} {\bibfnamefont {K.}~\bibnamefont
  {Abe}} \emph {et~al.} (\bibinfo {collaboration} {T2K}),\ }\href {\doibase
  10.1103/PhysRevD.97.032002} {\bibfield  {journal} {\bibinfo  {journal} {Phys.
  Rev. D}\ }\textbf {\bibinfo {volume} {97}},\ \bibinfo {pages} {032002}
  (\bibinfo {year} {2018})},\ \Eprint {http://arxiv.org/abs/1704.07467}
  {arXiv:1704.07467 [hep-ex]} \BibitemShut {NoStop}%
\bibitem [{\citenamefont {Abratenko}\ \emph
  {et~al.}(2021{\natexlab{a}})\citenamefont {Abratenko} \emph
  {et~al.}}]{MicroBooNE:2021ccs}%
  \BibitemOpen
  \bibfield  {author} {\bibinfo {author} {\bibfnamefont {P.}~\bibnamefont
  {Abratenko}} \emph {et~al.} (\bibinfo {collaboration} {MicroBooNE}),\
  }\href@noop {} {\bibfield  {journal} {\bibinfo  {journal} {(accepted by Phys.
  Rev. D)}\ } (\bibinfo {year} {2021}{\natexlab{a}})},\ \Eprint
  {http://arxiv.org/abs/2110.14028} {arXiv:2110.14028 [hep-ex]} \BibitemShut
  {NoStop}%
\bibitem [{\citenamefont {Andreopoulos}\ \emph {et~al.}(2010)\citenamefont
  {Andreopoulos} \emph {et~al.}}]{Andreopoulos:2009rq}%
  \BibitemOpen
  \bibfield  {author} {\bibinfo {author} {\bibfnamefont {C.}~\bibnamefont
  {Andreopoulos}} \emph {et~al.},\ }\href {\doibase 10.1016/j.nima.2009.12.009}
  {\bibfield  {journal} {\bibinfo  {journal} {Nucl. Instr. and Meth. A}\
  }\textbf {\bibinfo {volume} {614}},\ \bibinfo {pages} {87} (\bibinfo {year}
  {2010})},\ \Eprint {http://arxiv.org/abs/0905.2517} {arXiv:0905.2517
  [hep-ph]} \BibitemShut {NoStop}%
\bibitem [{\citenamefont {Tena-Vidal}\ \emph {et~al.}(2021)\citenamefont
  {Tena-Vidal} \emph {et~al.}}]{GENIE:2021zuu}%
  \BibitemOpen
  \bibfield  {author} {\bibinfo {author} {\bibfnamefont {J.}~\bibnamefont
  {Tena-Vidal}} \emph {et~al.} (\bibinfo {collaboration} {GENIE}),\ }\href
  {\doibase 10.1103/PhysRevD.104.072009} {\bibfield  {journal} {\bibinfo
  {journal} {Phys. Rev. D}\ }\textbf {\bibinfo {volume} {104}},\ \bibinfo
  {pages} {072009} (\bibinfo {year} {2021})},\ \Eprint
  {http://arxiv.org/abs/2104.09179} {arXiv:2104.09179 [hep-ph]} \BibitemShut
  {NoStop}%
\bibitem [{\citenamefont {Rein}\ and\ \citenamefont
  {Sehgal}(1981)}]{Rein:1980wg}%
  \BibitemOpen
  \bibfield  {author} {\bibinfo {author} {\bibfnamefont {D.}~\bibnamefont
  {Rein}}\ and\ \bibinfo {author} {\bibfnamefont {L.~M.}\ \bibnamefont
  {Sehgal}},\ }\href {\doibase 10.1016/0003-4916(81)90242-6} {\bibfield
  {journal} {\bibinfo  {journal} {Annals Phys.}\ }\textbf {\bibinfo {volume}
  {133}},\ \bibinfo {pages} {79} (\bibinfo {year} {1981})}\BibitemShut
  {NoStop}%
\bibitem [{\citenamefont {Berger}\ and\ \citenamefont
  {Sehgal}(2007)}]{Berger:2007rq}%
  \BibitemOpen
  \bibfield  {author} {\bibinfo {author} {\bibfnamefont {C.}~\bibnamefont
  {Berger}}\ and\ \bibinfo {author} {\bibfnamefont {L.~M.}\ \bibnamefont
  {Sehgal}},\ }\href {\doibase 10.1103/PhysRevD.76.113004} {\bibfield
  {journal} {\bibinfo  {journal} {Phys. Rev. D}\ }\textbf {\bibinfo {volume}
  {76}},\ \bibinfo {pages} {113004} (\bibinfo {year} {2007})},\ \Eprint
  {http://arxiv.org/abs/0709.4378} {arXiv:0709.4378 [hep-ph]} \BibitemShut
  {NoStop}%
\bibitem [{\citenamefont {Graczyk}\ and\ \citenamefont
  {Sobczyk}(2008)}]{Graczyk:2008}%
  \BibitemOpen
  \bibfield  {author} {\bibinfo {author} {\bibfnamefont {K.~M.}\ \bibnamefont
  {Graczyk}}\ and\ \bibinfo {author} {\bibfnamefont {J.~T.}\ \bibnamefont
  {Sobczyk}},\ }\href {\doibase 10.1103/physrevd.77.053003} {\bibfield
  {journal} {\bibinfo  {journal} {Physical Review D}\ }\textbf {\bibinfo
  {volume} {77}} (\bibinfo {year} {2008}),\
  10.1103/physrevd.77.053003}\BibitemShut {NoStop}%
\bibitem [{\citenamefont {Kuzmin}\ \emph {et~al.}(2004)\citenamefont {Kuzmin},
  \citenamefont {Lyubushkin},\ and\ \citenamefont {Naumov}}]{Kuzmin:2004}%
  \BibitemOpen
  \bibfield  {author} {\bibinfo {author} {\bibfnamefont {K.~S.}\ \bibnamefont
  {Kuzmin}}, \bibinfo {author} {\bibfnamefont {V.~V.}\ \bibnamefont
  {Lyubushkin}}, \ and\ \bibinfo {author} {\bibfnamefont {V.~A.}\ \bibnamefont
  {Naumov}},\ }\href {\doibase 10.1142/s0217732304016172} {\bibfield  {journal}
  {\bibinfo  {journal} {Modern Physics Letters A}\ }\textbf {\bibinfo {volume}
  {19}},\ \bibinfo {pages} {2815–2829} (\bibinfo {year} {2004})}\BibitemShut
  {NoStop}%
\bibitem [{\citenamefont {Nowak}(2009)}]{Nowak:2009}%
  \BibitemOpen
  \bibfield  {author} {\bibinfo {author} {\bibfnamefont {J.~A.}\ \bibnamefont
  {Nowak}},\ }\href {\doibase 10.1063/1.3274164} {\bibfield  {journal}
  {\bibinfo  {journal} {AIP Conference Proceedings}\ } (\bibinfo {year}
  {2009}),\ 10.1063/1.3274164}\BibitemShut {NoStop}%
\bibitem [{\citenamefont {Agostinelli}\ \emph {et~al.}(2003)\citenamefont
  {Agostinelli} \emph {et~al.}}]{GEANT4:2002zbu}%
  \BibitemOpen
  \bibfield  {author} {\bibinfo {author} {\bibfnamefont {S.}~\bibnamefont
  {Agostinelli}} \emph {et~al.} (\bibinfo {collaboration} {GEANT4}),\ }\href
  {\doibase 10.1016/S0168-9002(03)01368-8} {\bibfield  {journal} {\bibinfo
  {journal} {Nucl. Instrum. Meth. A}\ }\textbf {\bibinfo {volume} {506}},\
  \bibinfo {pages} {250} (\bibinfo {year} {2003})}\BibitemShut {NoStop}%
\bibitem [{\citenamefont {Snider}\ and\ \citenamefont
  {Petrillo}(2017)}]{Snider_2017}%
  \BibitemOpen
  \bibfield  {author} {\bibinfo {author} {\bibfnamefont {E.}~\bibnamefont
  {Snider}}\ and\ \bibinfo {author} {\bibfnamefont {G.}~\bibnamefont
  {Petrillo}},\ }\href {\doibase 10.1088/1742-6596/898/4/042057} {\bibfield
  {journal} {\bibinfo  {journal} {Journal of Physics: Conference Series}\
  }\textbf {\bibinfo {volume} {898}},\ \bibinfo {pages} {042057} (\bibinfo
  {year} {2017})}\BibitemShut {NoStop}%
\bibitem [{\citenamefont {Acciarri}\ \emph
  {et~al.}(2017{\natexlab{c}})\citenamefont {Acciarri} \emph
  {et~al.}}]{MicroBooNE:2017qiu}%
  \BibitemOpen
  \bibfield  {author} {\bibinfo {author} {\bibfnamefont {R.}~\bibnamefont
  {Acciarri}} \emph {et~al.} (\bibinfo {collaboration} {MicroBooNE}),\ }\href
  {\doibase 10.1088/1748-0221/12/08/P08003} {\bibfield  {journal} {\bibinfo
  {journal} {JINST}\ }\textbf {\bibinfo {volume} {12}},\ \bibinfo {pages}
  {P08003} (\bibinfo {year} {2017}{\natexlab{c}})},\ \Eprint
  {http://arxiv.org/abs/1705.07341} {arXiv:1705.07341 [physics.ins-det]}
  \BibitemShut {NoStop}%
\bibitem [{\citenamefont {Adams}\ \emph
  {et~al.}(2018{\natexlab{a}})\citenamefont {Adams} \emph
  {et~al.}}]{MicroBooNE:2018swd}%
  \BibitemOpen
  \bibfield  {author} {\bibinfo {author} {\bibfnamefont {C.}~\bibnamefont
  {Adams}} \emph {et~al.} (\bibinfo {collaboration} {MicroBooNE}),\ }\href
  {\doibase 10.1088/1748-0221/13/07/P07006} {\bibfield  {journal} {\bibinfo
  {journal} {JINST}\ }\textbf {\bibinfo {volume} {13}},\ \bibinfo {pages}
  {P07006} (\bibinfo {year} {2018}{\natexlab{a}})},\ \Eprint
  {http://arxiv.org/abs/1802.08709} {arXiv:1802.08709 [physics.ins-det]}
  \BibitemShut {NoStop}%
\bibitem [{\citenamefont {Adams}\ \emph
  {et~al.}(2018{\natexlab{b}})\citenamefont {Adams} \emph
  {et~al.}}]{MicroBooNE:2018vro}%
  \BibitemOpen
  \bibfield  {author} {\bibinfo {author} {\bibfnamefont {C.}~\bibnamefont
  {Adams}} \emph {et~al.} (\bibinfo {collaboration} {MicroBooNE}),\ }\href
  {\doibase 10.1088/1748-0221/13/07/P07007} {\bibfield  {journal} {\bibinfo
  {journal} {JINST}\ }\textbf {\bibinfo {volume} {13}},\ \bibinfo {pages}
  {P07007} (\bibinfo {year} {2018}{\natexlab{b}})},\ \Eprint
  {http://arxiv.org/abs/1804.02583} {arXiv:1804.02583 [physics.ins-det]}
  \BibitemShut {NoStop}%
\bibitem [{\citenamefont {Marshall}\ and\ \citenamefont
  {Thomson}(2015)}]{Marshall:2015rfa}%
  \BibitemOpen
  \bibfield  {author} {\bibinfo {author} {\bibfnamefont {J.~S.}\ \bibnamefont
  {Marshall}}\ and\ \bibinfo {author} {\bibfnamefont {M.~A.}\ \bibnamefont
  {Thomson}},\ }\href {\doibase 10.1140/epjc/s10052-015-3659-3} {\bibfield
  {journal} {\bibinfo  {journal} {Eur. Phys. J.}\ }\textbf {\bibinfo {volume}
  {C75}},\ \bibinfo {pages} {439} (\bibinfo {year} {2015})}\BibitemShut
  {NoStop}%
\bibitem [{\citenamefont {Chen}\ and\ \citenamefont
  {Guestrin}(2016)}]{Chen:2016:XST:2939672.2939785}%
  \BibitemOpen
  \bibfield  {author} {\bibinfo {author} {\bibfnamefont {T.}~\bibnamefont
  {Chen}}\ and\ \bibinfo {author} {\bibfnamefont {C.}~\bibnamefont
  {Guestrin}},\ }in\ \href {\doibase 10.1145/2939672.2939785} {\emph {\bibinfo
  {booktitle} {Proceedings of the 22nd ACM SIGKDD International Conference on
  Knowledge Discovery and Data Mining}}},\ \bibinfo {series and number} {KDD
  '16}\ (\bibinfo  {publisher} {ACM},\ \bibinfo {address} {New York, NY, USA},\
  \bibinfo {year} {2016})\ pp.\ \bibinfo {pages} {785--794}\BibitemShut
  {NoStop}%
\bibitem [{\citenamefont {Abratenko}\ \emph {et~al.}(2019)\citenamefont
  {Abratenko} \emph {et~al.}}]{MicroBooNE:2019nio}%
  \BibitemOpen
  \bibfield  {author} {\bibinfo {author} {\bibfnamefont {P.}~\bibnamefont
  {Abratenko}} \emph {et~al.} (\bibinfo {collaboration} {MicroBooNE}),\ }\href
  {\doibase 10.1103/PhysRevLett.123.131801} {\bibfield  {journal} {\bibinfo
  {journal} {Phys. Rev. Lett.}\ }\textbf {\bibinfo {volume} {123}},\ \bibinfo
  {pages} {131801} (\bibinfo {year} {2019})},\ \Eprint
  {http://arxiv.org/abs/1905.09694} {arXiv:1905.09694 [hep-ex]} \BibitemShut
  {NoStop}%
\bibitem [{\citenamefont {Calcutt}\ \emph {et~al.}(2021)\citenamefont
  {Calcutt}, \citenamefont {Thorpe}, \citenamefont {Mahn},\ and\ \citenamefont
  {Fields}}]{Calcutt:2021zck}%
  \BibitemOpen
  \bibfield  {author} {\bibinfo {author} {\bibfnamefont {J.}~\bibnamefont
  {Calcutt}}, \bibinfo {author} {\bibfnamefont {C.}~\bibnamefont {Thorpe}},
  \bibinfo {author} {\bibfnamefont {K.}~\bibnamefont {Mahn}}, \ and\ \bibinfo
  {author} {\bibfnamefont {L.}~\bibnamefont {Fields}},\ }\href {\doibase
  10.1088/1748-0221/16/08/P08042} {\bibfield  {journal} {\bibinfo  {journal}
  {JINST}\ }\textbf {\bibinfo {volume} {16}},\ \bibinfo {pages} {P08042}
  (\bibinfo {year} {2021})},\ \Eprint {http://arxiv.org/abs/2105.01744}
  {arXiv:2105.01744 [physics.data-an]} \BibitemShut {NoStop}%
\bibitem [{\citenamefont {Abratenko}\ \emph
  {et~al.}(2021{\natexlab{b}})\citenamefont {Abratenko} \emph
  {et~al.}}]{MicroBooNE:2021roa}%
  \BibitemOpen
  \bibfield  {author} {\bibinfo {author} {\bibfnamefont {P.}~\bibnamefont
  {Abratenko}} \emph {et~al.} (\bibinfo {collaboration} {MicroBooNE}),\
  }\href@noop {} {\  (\bibinfo {year} {2021}{\natexlab{b}})},\ \Eprint
  {http://arxiv.org/abs/2111.03556} {arXiv:2111.03556 [hep-ex]} \BibitemShut
  {NoStop}%
\bibitem [{\citenamefont {Adams}\ \emph {et~al.}(2020)\citenamefont {Adams}
  \emph {et~al.}}]{MicroBooNE:2019koz}%
  \BibitemOpen
  \bibfield  {author} {\bibinfo {author} {\bibfnamefont {C.}~\bibnamefont
  {Adams}} \emph {et~al.} (\bibinfo {collaboration} {MicroBooNE}),\ }\href
  {\doibase 10.1088/1748-0221/15/07/P07010} {\bibfield  {journal} {\bibinfo
  {journal} {JINST}\ }\textbf {\bibinfo {volume} {15}},\ \bibinfo {pages}
  {P07010} (\bibinfo {year} {2020})},\ \Eprint
  {http://arxiv.org/abs/1910.01430} {arXiv:1910.01430 [physics.ins-det]}
  \BibitemShut {NoStop}%
\bibitem [{\citenamefont {Abratenko}\ \emph {et~al.}(2020)\citenamefont
  {Abratenko} \emph {et~al.}}]{MicroBooNE:2020kca}%
  \BibitemOpen
  \bibfield  {author} {\bibinfo {author} {\bibfnamefont {P.}~\bibnamefont
  {Abratenko}} \emph {et~al.} (\bibinfo {collaboration} {MicroBooNE}),\ }\href
  {\doibase 10.1088/1748-0221/15/12/P12037} {\bibfield  {journal} {\bibinfo
  {journal} {JINST}\ }\textbf {\bibinfo {volume} {15}},\ \bibinfo {pages}
  {P12037} (\bibinfo {year} {2020})},\ \Eprint
  {http://arxiv.org/abs/2008.09765} {arXiv:2008.09765 [physics.ins-det]}
  \BibitemShut {NoStop}%
\bibitem [{\citenamefont {Caratelli}(2018)}]{caratelli2018}%
  \BibitemOpen
  \bibfield  {author} {\bibinfo {author} {\bibfnamefont {D.}~\bibnamefont
  {Caratelli}},\ }\emph {\bibinfo {title} {Study of Electromagnetic
  Interactions in the MicroBooNE Liquid Argon Time Projection Chamber}},\
  \href@noop {} {Ph.D. thesis},\ \bibinfo  {school} {Columbia University}
  (\bibinfo {year} {2018})\BibitemShut {NoStop}%
\bibitem [{\citenamefont {Zyla}\ \emph {et~al.}(2020)\citenamefont {Zyla} \emph
  {et~al.}}]{ParticleDataGroup:2020ssz}%
  \BibitemOpen
  \bibfield  {author} {\bibinfo {author} {\bibfnamefont {P.~A.}\ \bibnamefont
  {Zyla}} \emph {et~al.} (\bibinfo {collaboration} {Particle Data Group}),\
  }\href {\doibase 10.1093/ptep/ptaa104} {\bibfield  {journal} {\bibinfo
  {journal} {PTEP}\ }\textbf {\bibinfo {volume} {2020}},\ \bibinfo {pages}
  {083C01} (\bibinfo {year} {2020})}\BibitemShut {NoStop}%
\bibitem [{\citenamefont {Ji}\ \emph {et~al.}(2020)\citenamefont {Ji},
  \citenamefont {Gu}, \citenamefont {Qian}, \citenamefont {Wei},\ and\
  \citenamefont {Zhang}}]{Ji:2019yca}%
  \BibitemOpen
  \bibfield  {author} {\bibinfo {author} {\bibfnamefont {X.}~\bibnamefont
  {Ji}}, \bibinfo {author} {\bibfnamefont {W.}~\bibnamefont {Gu}}, \bibinfo
  {author} {\bibfnamefont {X.}~\bibnamefont {Qian}}, \bibinfo {author}
  {\bibfnamefont {H.}~\bibnamefont {Wei}}, \ and\ \bibinfo {author}
  {\bibfnamefont {C.}~\bibnamefont {Zhang}},\ }\href {\doibase
  10.1016/j.nima.2020.163677} {\bibfield  {journal} {\bibinfo  {journal} {Nucl.
  Instr. and Meth. A}\ }\textbf {\bibinfo {volume} {961}},\ \bibinfo {pages}
  {163677} (\bibinfo {year} {2020})},\ \Eprint
  {http://arxiv.org/abs/1903.07185} {arXiv:1903.07185 [physics.data-an]}
  \BibitemShut {NoStop}%
\bibitem [{\citenamefont {Messerly}\ \emph {et~al.}(2021)\citenamefont
  {Messerly} \emph {et~al.}}]{MINERvA:2021ddh}%
  \BibitemOpen
  \bibfield  {author} {\bibinfo {author} {\bibfnamefont {B.}~\bibnamefont
  {Messerly}} \emph {et~al.} (\bibinfo {collaboration} {MINERvA}),\ }\href
  {\doibase 10.1051/epjconf/202125103046} {\bibfield  {journal} {\bibinfo
  {journal} {EPJ Web Conf.}\ }\textbf {\bibinfo {volume} {251}},\ \bibinfo
  {pages} {03046} (\bibinfo {year} {2021})},\ \Eprint
  {http://arxiv.org/abs/2103.08677} {arXiv:2103.08677 [hep-ex]} \BibitemShut
  {NoStop}%
\bibitem [{\citenamefont {Golan}\ \emph {et~al.}(2012)\citenamefont {Golan},
  \citenamefont {Sobczyk},\ and\ \citenamefont {Zmuda}}]{Golan:2012rfa}%
  \BibitemOpen
  \bibfield  {author} {\bibinfo {author} {\bibfnamefont {T.}~\bibnamefont
  {Golan}}, \bibinfo {author} {\bibfnamefont {J.~T.}\ \bibnamefont {Sobczyk}},
  \ and\ \bibinfo {author} {\bibfnamefont {J.}~\bibnamefont {Zmuda}},\ }\href
  {\doibase 10.1016/j.nuclphysbps.2012.09.136} {\bibfield  {journal} {\bibinfo
  {journal} {Nucl. Phys. B Proc. Suppl.}\ }\textbf {\bibinfo {volume}
  {229-232}},\ \bibinfo {pages} {499} (\bibinfo {year} {2012})}\BibitemShut
  {NoStop}%
\bibitem [{\citenamefont {Hayato}(2002)}]{Hayato:2002sd}%
  \BibitemOpen
  \bibfield  {author} {\bibinfo {author} {\bibfnamefont {Y.}~\bibnamefont
  {Hayato}},\ }\href {\doibase 10.1016/S0920-5632(02)01759-0} {\bibfield
  {journal} {\bibinfo  {journal} {Nucl. Phys. B Proc. Suppl.}\ }\textbf
  {\bibinfo {volume} {112}},\ \bibinfo {pages} {171} (\bibinfo {year}
  {2002})}\BibitemShut {NoStop}%
\bibitem [{\citenamefont {Avanzini}\ \emph {et~al.}(2022)\citenamefont
  {Avanzini} \emph {et~al.}}]{Avanzini:2021qlx}%
  \BibitemOpen
  \bibfield  {author} {\bibinfo {author} {\bibfnamefont {M.~B.}\ \bibnamefont
  {Avanzini}} \emph {et~al.},\ }\href {\doibase 10.1103/PhysRevD.105.092004}
  {\bibfield  {journal} {\bibinfo  {journal} {Phys. Rev. D}\ }\textbf {\bibinfo
  {volume} {105}},\ \bibinfo {pages} {092004} (\bibinfo {year} {2022})},\
  \Eprint {http://arxiv.org/abs/2112.09194} {arXiv:2112.09194 [hep-ex]}
  \BibitemShut {NoStop}%
\bibitem [{\citenamefont {Casper}(2002)}]{Casper:2002sd}%
  \BibitemOpen
  \bibfield  {author} {\bibinfo {author} {\bibfnamefont {D.}~\bibnamefont
  {Casper}},\ }\href {\doibase 10.1016/S0920-5632(02)01756-5} {\bibfield
  {journal} {\bibinfo  {journal} {Nucl. Phys. B Proc. Suppl.}\ }\textbf
  {\bibinfo {volume} {112}},\ \bibinfo {pages} {161} (\bibinfo {year}
  {2002})},\ \Eprint {http://arxiv.org/abs/hep-ph/0208030}
  {arXiv:hep-ph/0208030} \BibitemShut {NoStop}%
\end{thebibliography}%
\nocite{*}
\end{document}